\documentclass[aps,prd,
nofootinbib,
preprintnumbers,
twocolumn,
showpacs, 
floatfix
]{revtex4-1}
\usepackage{scalefnt}
\usepackage{amsmath}
\usepackage{amssymb}
\usepackage{epsfig}
\usepackage{graphics}
\usepackage{color}
\usepackage[normal]{subfigure}
\usepackage{rotating}
\newcommand{\ewt}{\end{widetext}}
\newcommand{\be}{\begin{equation}}
\newcommand{\ee}{\end{equation}}
\newcommand{\bdm}{\begin{displaymath}}
\newcommand{\edm}{\end{displaymath}}
\newcommand{\bea}{\begin{eqnarray}}
\newcommand{\eea}{\end{eqnarray}}
\newcommand{\nn}{\nonumber}

\newcommand{\abbrev}{\scalefont{.9}}
\newcommand{\msbar}{$\overline{\mbox{\abbrev MS}}$}
\newcommand{\drbar}{$\overline{\mbox{\abbrev DR}}$}
\def\eq#1{{Eq.~(\ref{#1})}}
\def\eqs#1#2{{Eqs.~(\ref{#1})--(\ref{#2})}}
\def\fig#1{{Fig.~\ref{#1}}}

\def\Table#1{{Table~\ref{#1}}}

\def\sect#1{{Section~\ref{#1}}}

\def\app#1{{Appendix~\ref{#1}}}

\def\vev#1{\left\langle #1 \right\rangle}

\def\Tr{\mbox{Tr}\,}

\begin{document}
\preprint{TTP13-010, SFB/CPP-13-24}
%----------------------------------------------------------------------------------
\title{Unification scale vs.~electroweak-triplet mass in the SU(5)+${\bf 24_F}$ model at three loops}
%\pacs{11.25.Db 11.30.Pb 12.10.Kt 12.38.Bx}
\pacs{12.10.-g 11.15.Bt}
%----------------------------------------------------------------------------------
\author{Luca Di Luzio}\email{diluzio@kit.edu}
\author{Luminita Mihaila}\email{luminita.mihaila@kit.edu}
\affiliation{Institut f\"{u}r Theoretische Teilchenphysik,
Karlsruhe Institute of Technology (KIT), D-76128 Karlsruhe, Germany}
%----------------------------------------------------------------------------------
\begin{abstract}
It was shown recently that the original SU(5) theory of Georgi and Glashow,
augmented with an adjoint fermionic multiplet $24_F$, can be made compatible both with neutrino masses
and gauge coupling unification. In particular, the model predicts that 
either electroweak-triplet states are light, within the reach of the Large Hadron Collider 
(LHC), or proton decay will become accessible at the next generation of megaton-scale facilities.
In this paper, we present the computation of the correlation function 
between the electroweak-triplet masses and the unification scale at the next-to-next-to-leading-order (NNLO). 
Such an accuracy on the theory side is necessary in order to settle the convergence of the
perturbative expansion and to match the experimental precision on the
determination of the electroweak gauge couplings at the Z-boson mass scale.
\end{abstract}
\maketitle
%----------------------------------------------------------------------------------

%%%%%%%%%%%%%%%%%%%%%%%%%%%%%%%%%%%%%%%%%%%
\section{Introduction}
\label{intro}
%%%%%%%%%%%%%%%%%%%%%%%%%%%%%%%%%%%%%%%%%%%

The quantum numbers of the Standard Model (SM) fermions together with the apparent convergence 
of the strong and electroweak couplings at energies below the Planck scale 
point towards a unified description of the SM interactions. 
One of the fundamental predictions of a Grand Unified Theory (GUT) is the existence 
of baryon and lepton number violating interactions which can manifest themselves 
at low energy via matter instability  
(for a review see for example Ref.~\cite{Nath:2006ut}). 
Though the decay of the proton has not been observed so far, the lower bound on
the proton lifetime, together with the low-energy values of the SM gauge couplings 
and the SM fermion masses and mixings provide us severe constraints
on the class of viable GUT models.

On the other hand, the degree of complexity of GUTs, even in their simplest realizations,
makes them hard to be tested. It is enough to say that one of the 
few absolute certainties about grand unification today is that the original 
SU(5) model of Georgi and Glashow (GG) \cite{Georgi:1974sy} 
is ruled out. In particular, the failure of the minimal 
model can be attributed 
both to the lack of gauge coupling unification \cite{Ellis:1990wk,Langacker:1991an,Amaldi:1991cn}
and to the fact that an accidental $B-L$ global symmetry \cite{Wilczek:1979et}, 
as in the SM, prevents neutrinos to be massive. 

When looking for a minimal realistic extension of the GG model 
it would be economical (and hence predictive) if the solution 
to the issue of gauge coupling unification were related to the generation of neutrino masses. 
This is, essentially, the philosophy behind two recent proposals where 
an extra scalar representation $15_H$ \cite{Dorsner:2005fq,*Dorsner:2005ii}, 
or alternatively, a fermionic representation $24_F$ \cite{Bajc:2006ia,*Bajc:2007zf} 
is added to the field content of the model. 
In both cases, the extra degrees of freedom have the right quantum numbers to generate neutrino masses via the seesaw 
mechanism \cite{Minkowski:1977sc,*GellMann:1980vs,*Yanagida:1979as,*Glashow:1979nm,*Mohapatra:1979ia,Magg:1980ut,*Schechter:1980gr,*Lazarides:1980nt,*Mohapatra:1980yp,Foot:1988aq} and restore unification by properly modifying the running of the gauge couplings.  
Though both the models share a similar degree of minimality, 
we shall restrict our discussion to the ${\rm SU(5)}+24_F$ model and postpone the
${\rm SU(5)}+15_H$ model for a future investigation.

Let us briefly recall the reason why gauge coupling unification fails within 
the minimal GG model. While $\alpha_2$ and $\alpha_3$ meet around $10^{16}$ GeV, 
the main issue is the early convergence of $\alpha_2$ and $\alpha_1$
at about $10^{13}$~GeV~\cite{Ellis:1990wk,Langacker:1991an,Amaldi:1991cn}, at
odds with the bounds 
enforced by the nonobservation of the proton 
decay. More precisely, assuming no cancellations in the flavour structure of the 
gauge-induced proton decay rates \cite{FileviezPerez:2004hn,Dorsner:2004xa}, the latest experimental 
data from the Super-Kamiokande observatory~\cite{Nishino:2012ipa} 
imply a conservative lower bound on the unification scale $M_G$ of about
$10^{15.5} \ \rm{GeV}$.
Hence, the key ingredients for a viable unification pattern are
additional particles 
charged under the ${\rm SU(2)}_L$ group that  delay the meeting of $\alpha_1$ and $\alpha_2$. 
Such a role in the ${\rm SU(5)}+24_F$ model can be only played by the 
electroweak fermion and scalar triplets $(1,3,0)_{F,H} \in 24_{F,H}$, living in the $24$-dimensional
representations of the SU(5) gauge group. They 
are predicted to be light \cite{Bajc:2006ia,*Bajc:2007zf}, eventually of $\mathcal{O}(\text{TeV})$, 
so that a large enough unification scale can be reached. \\
Both types of  triplets, if light enough, can give interesting signature
at the LHC.  
The fermionic component leads to same sign dilepton events which violate lepton number \cite{Bajc:2006ia,*Bajc:2007zf}
(see \cite{Franceschini:2008pz,delAguila:2008cj,delAguila:2008hw,Arhrib:2009mz,ATLAS1} for some recent collider 
analysis). The bosonic triplet instead can easily modify 
the decay properties of the Higgs boson (see e.g.~\cite{Chang:2012ta}), 
that will be measured with increasingly precision at the LHC. 

The complete unification pattern including also 
the convergence of $\alpha_3$ with $\alpha_{1}$ and $\alpha_{2}$ requires heavier 
particles charged under the ${\rm SU(3)}_C$ group. In the ${\rm SU(5)}+24_F$ model these are  
the colour-octet fermions and scalars, $(8,1,0)_{F,H} \in 24_{F,H}$,
that are predicted to live 
at intermediate mass scales of about
$10^{8}$GeV~\cite{Bajc:2006ia,*Bajc:2007zf}, well beyond the LHC energy
range. 

Remarkably, it can be established a correlation between the electroweak
triplet masses and the unification scale which acts as a ``precision observable''. 
Imposing  the condition of gauge coupling unification, the electroweak
triplet masses  can be expressed through the Renormalization Group
Equations (RGEs) of the model as a function of  the GUT scale and
the electroweak couplings $\alpha_1$ and $\alpha_2$ evaluated 
at the Z-boson mass scale $M_Z$. Given the high accuracy at which the
latter parameters are determined experimentally, one can make very
precise predictions for the dependence of the  electroweak
triplet masses on the GUT scale. Such a correlation function plays a
significant role for testing the 
model. If the electroweak triplets are not 
found at the LHC, then, according to the ${\rm SU(5)}+24_F$ model, the predicted unification scale 
is smaller than about $10^{16}$~GeV. Thus, matter instability is
expected to be observed in the next generation of  
proton decay experiments \cite{Abe:2011ts}, otherwise the model is ruled out. For such an
important task it is mandatory to have precise theoretical predictions
at least comparable with the experimental accuracy. 
Let us also mention that the magnitude of the two-loop radiative
corrections~\cite{Bajc:2007zf} to the determination of the triplet masses  
 is comparable with that of the one-loop
contributions and it is almost 10 times larger than the
parametric uncertainty due to the dependence on the low-energy values of $\alpha_1$ and $\alpha_2$.
Thus, a three-loop
analysis is indispensable in order to establish whether the perturbative
series converges and to match the experimental precision.

For a consistent three-loop prediction of the electroweak-triplet masses, one
needs the RGEs of the gauge couplings
for the ${\rm SU(5)}+24_F$ model and for all effective field theories
(including the SM) that can be derived from it, at three-loop
accuracy. In addition, threshold corrections induced at the heavy
particle mass scales are necessary at the two-loop order.
The RGEs for gauge theories based on semisimple gauge groups have been known 
at two-loop accuracy for a long time~\cite{Machacek:1983tz,Jack:1984vj}, whereas for simple gauge groups even the
three-loop order results are known~\cite{Pickering:2001aq}. The three-loop contributions
to the RGEs of the SM~\cite{Mihaila:2012fm,Bednyakov:2012en,Chetyrkin:2012rz} have been
computed recently. 
 In this work we go a step further towards the computation of the
three-loop corrections to the RGEs for a general semisimple gauge
group. \\
The threshold corrections for a general gauge theory are known at the one-loop level also since
long time~\cite{Weinberg:1980wa,*Hall:1980kf}. However, general results for the two-loop
contributions are not available in the literature. In this paper we
also compute the two-loop threshold corrections for the relevant heavy states of the ${\rm SU(5)}+24_F$ model.

The paper is organised as follows: in the next Section we introduce the
${\rm SU(5)}+24_F$ model, specify the particle content and 
describe its main features. In~\sect{let} and \sect{motiv3loop} we discuss the approach of
multi-loop calculations within  effective field theories using mass
independent regularization and renormalization schemes. Furthermore, we 
present our results for   the three-loop gauge  beta functions and the two-loop
matching coefficients for the effective theory consisting in the SM
and electroweak triplets. The corresponding results for the effective
theory including also colour-octet multiplets are given in~\app{Ocontmatch}. In~\sect{numan}, we describe the
phenomenological implications of our calculation. Especially, we
emphasize the effects of the three-loop corrections on the prediction of
the electroweak-triplet masses. Finally in~\sect{concl}
we present our conclusions and insights. In addition, we discuss in
some detail, in~\app{details24F}, the tree-level calculation of
the mass spectrum of the ${\rm SU(5)}+24_F$ model and the relations
that can be established between its parameters and the ones occurring in
the low-energy effective theory.

%%%%%%%%%%%%%%%%%%%%%%%%%%%%%%%%%%%%%%%%%%%
\section{The SU(5) + ${\bf 24_F}$ model}
\label{SU524Fmodel}
%%%%%%%%%%%%%%%%%%%%%%%%%%%%%%%%%%%%%%%%%%%

Let us start by reviewing the basic features of the 
SU(5) model augmented with a fermionic $24_F$ multiplet.   
More technical aspects about the particle content, its mass spectrum
and low-energy interactions are deferred into a self-contained 
Appendix (cf.~\app{details24F}). 

The scalar sector spans over two different representations, namely, 
\be
5_H = \underbrace{(3,1,-\tfrac{1}{3})_H}_{\mathcal{T}} \oplus \underbrace{(1,2,+\tfrac{1}{2})_H}_h
\, ,
\ee
and
\begin{multline}
24_H  = \underbrace{(1,1,0)_H}_{S_H} \oplus \underbrace{(1,3,0)_H}_{T_H}
\oplus \underbrace{(8,1,0)_H}_{O_H} \\ \oplus   
\underbrace{(3,2,-\tfrac{5}{6})_H}_{X_H} \oplus
\underbrace{(\overline{3},2,+\tfrac{5}{6})_H}_{\overline{X}_H}  
\, ,
\end{multline}
where $S_H$, $T_H$ and $O_H$ ($\mathcal{T}$, $h$ and $X_H$) are real (complex) scalars. 
In our notation, $h$ stands for the SM Higgs doublet. \\
The decomposition of the vector bosons belonging to the SU(5) adjoint representation 
reads
\begin{multline}
24_V = \underbrace{(1,1,0)_V}_{S_V} \oplus \underbrace{(1,3,0)_V}_{T_V}
\oplus \underbrace{(8,1,0)_V}_{O_V} \oplus \\ 
\underbrace{(3,2,-\tfrac{5}{6})_V}_{X_V} \oplus
\underbrace{(\overline{3},2,+\tfrac{5}{6})_V}_{\overline{X}_V}  
\, ,
\end{multline}
where $S_V$, $T_V$ and $O_V$ denote the SM gauge bosons,  
while $X_V$ and $\overline{X}_V$ correspond to the super-heavy gauge
bosons of the SU(5) broken phase, the so-called leptoquarks. They are 
responsible for the gauge-induced proton-decay rate 
and include, as a longitudinal component, 
the Goldstone boson $X_H$ of the SU(5) broken phase. \\
Finally, the matter content of the model is given by
the Weyl fermions of the three SM families
\begin{align}
\overline{5}_F &= (\underbrace{\overline{3},1,+\tfrac{1}{3})_F}_{d^c}
\oplus \underbrace{(1,2,-\tfrac{1}{2})_F}_{\ell}  
\, , \\
10_F &= (\underbrace{\overline{3},1,-\tfrac{2}{3})_F}_{u^c} \oplus
\underbrace{(3,2,+\tfrac{1}{6})_F}_{q} \oplus
\underbrace{(1,1,+1)_F}_{e^c} 
\, , 
\end{align}
and the additional fermionic multiplet
\begin{multline}
24_F = \underbrace{(1,1,0)_F}_{S_F} \oplus \underbrace{(1,3,0)_F}_{T_F}
\oplus \underbrace{(8,1,0)_F}_{O_F} \oplus \\ 
\underbrace{(3,2,-\tfrac{5}{6})_F}_{X_F} \oplus
\underbrace{(\overline{3},2,+\tfrac{5}{6})_F}_{\overline{X}_F}  
\, ,
\end{multline}
where $S_F$, $T_F$, $O_F$ ($X_F$) are Majorana (Dirac) degrees of freedom.
A special role in the model is played by the electroweak singlet and triplet states 
$S_F$ and $T_F$. They are involved in the Yukawa interactions that
after the SU(5) gauge-symmetry breaking will generate masses for
neutrinos through a hybrid type-I+III seesaw mechanism \cite{Bajc:2006ia,*Bajc:2007zf} (for details see~\app{Yuksect}).
The electroweak singlet $S_F$ resembles a sterile neutrino, whereas the electroweak 
triplet is sometimes referred to as a heavy lepton.

Let us mention at this point that, as in the original SU(5) model,
nonrenormalizable operators are required in order to reproduce 
fermion masses and mixing~\cite{Ellis:1979fg,Dorsner:2006hw}.
Furthermore, the Higgs sector is the one of the genuine SU(5) model 
and the minimization of the scalar potential proceeds as usual (for
details of the calculation see~\app{scalsect}). 
All the states are subject to the constraints coming 
from the calculation of the tree-level spectrum. 
In this respect, though the required mass hierarchy strengthen the
fine-tuning issue typical for GUTs,   
it is nevertheless a nontrivial fact that the tree-level calculation of the spectrum 
allows the mass pattern required by unification \cite{Bajc:2006ia,*Bajc:2007zf} 
(for more details see also~\app{massspect}).

%%%%%%%%%%%%%%%%%%%%%%%%%%%%%%%%%%%%%%%%%%%%%%%%%%%%

\section{Effective field theory approach}
\label{let}

In the following, we concentrate on the study of the gauge coupling unification assuming 
the mass hierarchy 
\begin{equation}
m_{T_F}\approx m_{T_H} \ll m_{O_F}\approx m_{O_H} \ll M_G \, . 
\end{equation}
For such a largely split mass spectrum, it is convenient to
apply the method of effective field theories (EFTs). This approach was
introduced a long time ago in the context of GUTs~\cite{Weinberg:1980wa,Hall:1980kf} 
and has been extensively applied in the context of the SM
and its supersymmetric extension even in high
precision calculations (see for example
Refs.~\cite{Schroder:2005hy,Chetyrkin:2005ia,Kurz:2012ff}).      
It consists in integrating out the heavy degrees of freedom that 
cannot influence the physics at the low-energy scale. \\
In physical renormalizations schemes like the momentum subtraction
scheme or the on-shell scheme, the effects due to heavy particle
thresholds are included in
the renormalization constants of the parameters. However, for the analysis of the gauge coupling
unification that requires the running of the couplings over many orders
of magnitude, higher order radiative corrections to the RGEs are
essential. Nevertheless, their calculation beyond 
one-loop order in mass dependent 
renormalization schemes is quite involved. A much more suited scheme for
this purpose is the minimal subtraction scheme
(\msbar)~\cite{Bardeen:1978yd}, for which 
the gauge coupling beta functions are mass independent and their computation is
substantially simplified. Nevertheless, in this scheme the
Appelquist-Carazzone~\cite{Appelquist:1974tg}  
theorem does not hold anymore and  
the threshold effects have to be taken into account explicitly. The
latter are parametrized through the 
decoupling (or matching) coefficients. They can be computed 
perturbatively using the physical constraint that the Green's functions involving light particles
have to be equal in the original and the effective theory. \\
For the computation presented in this paper, we adopt this second method
and apply it up to the  
third order in perturbation theory.

Because in the underlying theory we can identify three well-separated mass scales
corresponding to electroweak triplets ($T_{H,F}$), colour octets ($O_{H,F}$) and GUT
particles ($\mathcal{T}$, $X_F$ and $X_V$), it is natural to construct a series of three effective
theories to take into account the individual mass thresholds.
 A summary of the individual
ingredients of the calculation is given in
\Table{tab:run_dec}.
For the present analysis we computed the  following  missing pieces: (i)
The three-loop RGEs for the gauge couplings of the effective
theory obtained by integrating out the super-heavy (GUT) particles. We
denote this EFT as SM+T+O; (ii) The three-loop RGEs for the gauge couplings
of the EFT obtained by integrating out the GUT particles and the octet
multiplets, that we call SM+T. In principle, a fourth effective
theory can be obtained if the mass pattern of the super-heavy particles
is taken into account. Especially, the mass of the $X_F$ state from the
$24_F$ multiplet can be at most of the order of $M_G^2/\Lambda$ (cf.~\eq{mXFftTO}). 
Here, $\Lambda$ is the cutoff of the effective SU(5) theory which should be chosen 
so that the correct $m_b/m_{\tau}$ ratio is reproduced and
the perturbativity domain is maximized. For the purpose of comparison with 
Ref.~\cite{Bajc:2006ia} we take the value $\Lambda=100 \, M_G$, 
though also lower values of $\Lambda$ are in principle viable \cite{Dorsner:2006fx}. 
In particular, for the contribution of $X_F$ to the running within the SM+T+O+${\rm X_F}$ 
EFT, we employ only a two-loop
analysis~\cite{Machacek:1983tz}, since it has a subdominant effect.
Furthermore, we compute the contributions of the electroweak triplets
and colour octets (both bosonic and  
fermionic components) to the two-loop matching coefficients of the the
SM gauge couplings, while the GUT-scale thresholds are considered only at the one-loop 
level~\cite{Weinberg:1980wa,Hall:1980kf}.\footnote{For a recent attempt towards the 
calculation of two-loop matching at the GUT scale see Ref.~\cite{Martens:2010pe}.}

\begin{table*}[ht]
  \begin{center}  
      \begin{tabular}{c|cccc}
%{p{2.5cm}p{2.5cm}p{2.5cm}p{2.5cm}p{2.5cm}p{2.5cm}}
        \hline
        \hline
        {\bf Running} & $\rm{SM}$ & $\rm{SM+T}$ & $\rm{SM+T+O}$ & $\rm{SM+T+O+X_F}$ \\ 
        \hline
        Scale & $M_Z\to\mu_{T}$ & $\mu_{T}\to\mu_{O}$ & $\mu_{O} \rightarrow \mu_{X_F}$ & $\mu_{X_F} \rightarrow M_{G}$ \\ 
        \hline
        $\#$ of loops & 3 & \bf{3} & \bf{3} & 2(3) \\ 
        \hline
        \hline
        {\bf Matching} & $\alpha_i^{\rm{SM}}\to\alpha_i^{\rm{SM+T}}$ & $\alpha_i^{\rm{SM+T}}\to\alpha_i^{\rm{SM+T+O}}$ 
        & $\alpha_i^{\rm{SM+T+O}}\to\alpha_i^{\rm{SM+T+O+X_F}}$ 
        & $\alpha_i^{\rm{SM+T+O+X_F}}\to\alpha_G$ \\ 
        \hline
        Scale & $\mu_{T}$ & $\mu_{O}$ & $\mu_{X_F}$ & $M_G$ \\ 
        \hline
        $\#$ of loops & \bf{2} & \bf{2} & 1(2) & 1(2) \\ 
        \hline
        \hline
      \end{tabular}
    \caption{\label{tab:run_dec}Loop corrections available for the individual
      steps of the running and matching procedure in the ${\rm SU(5)}+24_F$ model. The numbers in bold face are 
      due to the computation performed in this work, while the numbers in parentheses
      indicate the last missing ingredient for a complete three-loop analysis.
      }
  \end{center}
\end{table*}

%\clearpage

%===========================================================
\section{Running and decoupling }
\label{motiv3loop}
%===========================================================

For exemplification, we describe in the following the calculation done in 
the effective theory obtained integrating out the GUT particles and the 
colour-octet multiplets. Thus the particle content of the effective
theory consists in the SM particles and the electroweak triplets.\\
Let us introduce at this point the framework of the calculation. 
The most general Lagrangian containing the
renormalizable interactions 
of the SM fields and the  ${\rm SU(2)}_L$ triplets 
$T_{H,F}$ is given by\footnote{Yukawa interactions between the fermionic triplets and the 
SM fields can be safely neglected, since for light $\mathcal{O}(\text{TeV})$ triplets the 
new Yukawa couplings $y_T$ are bounded to be small in order to reproduce neutrino 
masses (cf.~\eq{mmnu}).}
\begin{align}
\label{THFSMlag}
\mathcal{L} &= \mathcal{L}^{\rm{SM}} 
+ \tfrac{1}{2} \left| D_{\mu} T_H \right|^2 
+ \tfrac{1}{2} \overline{T}_F \, i \, \gamma^\mu D_{\mu} T_F - V^{\rm{ren}}(h,T_H) \nn \\
&+ \text{gauge fixing} + \text{ghosts} 
\, ,
\end{align}
where the covariant derivative is defined as 
\be
D_{\mu} = \partial_\mu - i g_2 \, T_{\rm{adj}}^A W^A_\mu \, ,
\ee
with $T_{\rm{adj}}^A$ the  generators  of the ${\rm SU(2)}_L$ gauge group in
the adjoint representation. They are related to the  structure constants
by the relation 
$(T_{\rm{adj}}^A)_{BC} \equiv - i f^{ABC}$.
The scalar potential describing the quartic interactions $V_{\rm{4sc}}^{\rm{ren}}$, including the
SM Higgs doublet $h$, reads  
\begin{equation}
\label{Vren}
V_{\rm{4sc}}^{\rm{ren}}(h,T_H) = 
\lambda_h \left|h\right|^4
+ \tfrac{\lambda_{T}}{2} \left|T_H\right|^2 
+ \lambda_{hT} \, \left|h\right|^2 \left|T_H\right|^2 \, ,
\end{equation}
where $\lambda_h$ is the SM quartic coupling and $\lambda_{T}$ and
$\lambda_{hT}$ are new couplings. The tree-level relations between these low-energy
couplings and the Lagrangian parameters of the ${\rm SU(5)}+24_F$ model are 
given in Eqs.~(\ref{lamT}), (\ref{lambdaH}) and (\ref{lambdaHT}) of \app{lowenint}.

For later convenience we introduce the relevant coupling constants in terms 
of which  the analytical results are presented: 
$\alpha_i = g_i^2 / (4 \pi)$ with 
$i=1,2,3$, are the gauge coupling constants
$\alpha_t = y_t^2/ (4 \pi)$ where $y_t$ is the top-Yukawa coupling, and
$\alpha_{\lambda_h} = \lambda_h/ (4 \pi)$,
$\alpha_{\lambda_T} = \lambda_T/ (4 \pi)$
and 
$\alpha_{\lambda_{hT}} = \lambda_{hT}/ (4 \pi)$ denote the quartic
coupling constants in the scalar sector.
In the calculation, we adopt  the SU(5)-like normalization of the
$\alpha_1$ coupling. The three gauge coupling constants are  related to the quantities
usually used in the SM by the all-order relations
\begin{eqnarray}
  \alpha_1 &=& \frac{5}{3}\frac{\alpha_{\rm QED}}{\cos^2\theta_W}\,,\nonumber\\
  \alpha_2 &=& \frac{\alpha_{\rm QED}}{\sin^2\theta_W}\,,\nonumber\\
  \alpha_3 &=& \alpha_s\,,
\end{eqnarray}
where $\alpha_{\rm QED}$ is the fine structure constant, $\theta_W$ 
stands for the weak mixing angle and $\alpha_s$ is the strong coupling
constant.
Let us stress that the gauge couplings that we need are those defined in the theory described by the
Lagrangian given in Eq.~(\ref{THFSMlag}). They can be related to the SM
parameters through  the decoupling coefficients that we present in the
next section.

Furthermore, all the group theoretical factors we encountered in the
three-loop order calculation
can be expressed in terms of quadratic Casimir invariants of the
relevant representations of the gauge group. 
For a field transforming under the representation $R$ of the gauge group $G$, 
where the generators $R^A$ satisfy
\be
\left[ R^A, R^B \right] = i f^{ABC} R^C \, ,
\ee
the Casimir invariants are defined as follows 
\be
\begin{gathered}
\Tr (R^A R^B) = \delta^{AB} T_R \, , \qquad 
R^A_{ac} R^A_{cb} = \delta_{ab} C_R \, , \\
f^{ACD} f^{BCD} = \delta^{AB} C_G \, , \qquad
\delta^{AA} = N_G \, .
\end{gathered}
\ee
Here, $N_G$ denotes the dimension of the group.
Then the following relation, $C_R N_R = T_R N_G$, 
where $N_R = \delta_{aa}$ is the dimension of representation $R$,
holds as well. 

In our case, the underlying gauge group is the same as the one of the SM, namely
${\rm SU(3)}_C\otimes {\rm SU(2)}_L\otimes {\rm U(1)}_Y$. To avoid confusion, we introduce
an additional index for the
Casimir invariants associated with the individual simple groups. Namely,
an index $C$ for the ${\rm SU(3)}_C$ group, an index $L$ for the ${\rm SU(2)}_L$
group and finally an index $Y$ for the ${\rm U(1)}_Y$ group. The explicit
notation and the numerical values can be found in \Table{tab:inv}.
The  numerical values for the hypercharges of the SM fermions and scalars
in the SU(5) normalization can be read from the discussion after \eq{deltab3T} and \eq{deltab3O}.
\begin{table}[ht]
%\centering
\begin{tabular}{|c|c|c|}
\hline
${\rm SU(3)}_C$ & ${\rm SU(2)}_L$ & $ {\rm U(1)}_Y$\\
\hline
\hline
$C_{G_c}=3$ & $C_{G_L}=2$ & $C_{G_Y}=0$\\
$C_{R_C}=\tfrac{4}{3}$ & $C_{R_L}=\tfrac{3}{4}$& $Y_{R_y}^2$ \\
$T_{R_C}=\tfrac{1}{2}$ & $T_{R_L}=\tfrac{1}{2}$& $Y_{R_y}^2$ \\
$N_{R_C}=3$ & $N_{R_L}=2$& $N_{R_Y}=1$ \\
$N_{G_C}=8$ & $N_{G_L}=3$& $N_{G_Y}=1 $
\\
\hline
\end{tabular}
\caption{Notations and numerical values for the Casimir invariants of the simple subgroups of 
${\rm SU(3)}_C\otimes {\rm SU(2)}_L\otimes {\rm U(1)}_Y$ that occur in the three-loop 
calculation. Here $R$ stands for the fundamental representations.}
\label{tab:inv}
\end{table}

%%%%%%%%%%%%%%%%%%%%%%%%%%%%%%%%%%%%%%%%%%%%%%%%%
%===========================================================
\subsection{Beta functions}
\label{betafunct}
%===========================================================
The energy dependence of the gauge couplings is controlled by the beta
functions. These are defined as 
\begin{eqnarray}
\label{defbf}
&& \mu^2 \frac{d}{d \mu^2} \frac{\alpha_i}{\pi} = \beta_i (\{\alpha_j\},\epsilon)
= -\epsilon \frac{\alpha_i}{\pi}  \\ 
&& - \left( \frac{\alpha_i}{\pi} \right)^2 
\left[ 
a_i + \sum_j \frac{\alpha_j}{\pi} b_{ij} 
  + \sum_{j,k} \frac{\alpha_j}{\pi} \frac{\alpha_k}{\pi} c_{ijk} + \ldots
\right] \, , \nn
\end{eqnarray}
with $i=1,2$ or $3$. The expression  after the second equality sign
gives the perturbative  expansion. Here, $\epsilon=(4-d)/2$ is the
regulator of Dimensional Regularization with $d$ being the space-time dimension used
for the evaluation of the momentum integrals.
 In practice, the functions $\beta_i$ are obtained from the renormalization
constants of the corresponding  couplings that are
defined as $\alpha_i^{\rm bare}=\mu^{2\epsilon} Z_{\alpha_i} \alpha_i$. Exploiting
the fact that the bare couplings 
 are $\mu$-independent and taking
into account that $ Z_{\alpha_i}$ may depend on all the other couplings
leads to the following formula:
\begin{equation}
  \label{eq::renconst_beta}
  \beta_i = 
  -\left[\epsilon\frac{\alpha_i}{\pi}
    +\frac{\alpha_i}{Z_{\alpha_i}}
    \sum_{{j},{j \neq i}}
    \frac{\partial Z_{\alpha_i}}{\partial \alpha_j}\beta_j\right]
  \left(1+\frac{\alpha_i}{Z_{\alpha_i}}
    \frac{\partial Z_{\alpha_i}}{\partial \alpha_i}\right)^{-1} \,,
\end{equation}
From Eq.~(\ref{eq::renconst_beta}) it is clear that the renormalization
constants $Z_{\alpha_i}\, (i=1,2,3)$ have to be computed up to
three-loop order. In principle each vertex containing the gauge coupling
$\alpha_i$ at tree level can be used in order to obtain $ Z_{\alpha_i}$
via the Slavnov-Taylor identity
\begin{eqnarray}
  Z_{\alpha_i} &=& \frac{(Z_{\text{vrtx}})^2}{\prod_k Z_{k,{\text{wf}}}}\,, 
  \label{eq::Zalpha}
\end{eqnarray}
where $Z_{\text{vrtx}}$ stands for the renormalization constant of the
vertex and $Z_{k,{\text{wf}}}$ for the wave function renormalization constant;
$k$ runs over all external particles. 

We have computed $Z_{\alpha_2}$ and $Z_{\alpha_3}$ using the
(Fadeev-Popov) ghost-gluon and the (Fadeev-Popov) ghost-$W$ vertices as they are the most
economical ones with respect to (wrt) number of diagrams. For $Z_{\alpha_1}$, a Ward identity
guarantees that there is a cancellation
between the vertex and some of the wave function
renormalization constants yielding 
\begin{eqnarray}
  Z_{\alpha_1} &=& \frac{1}{Z_Y},
\end{eqnarray}
where $Z_Y$ is the wave function renormalization constant 
for the  gauge boson of the ${\rm U(1)}_Y$ subgroup of the SM in 
the unbroken phase.

\begin{figure}[h]
\includegraphics[angle=0,width=8.5cm]{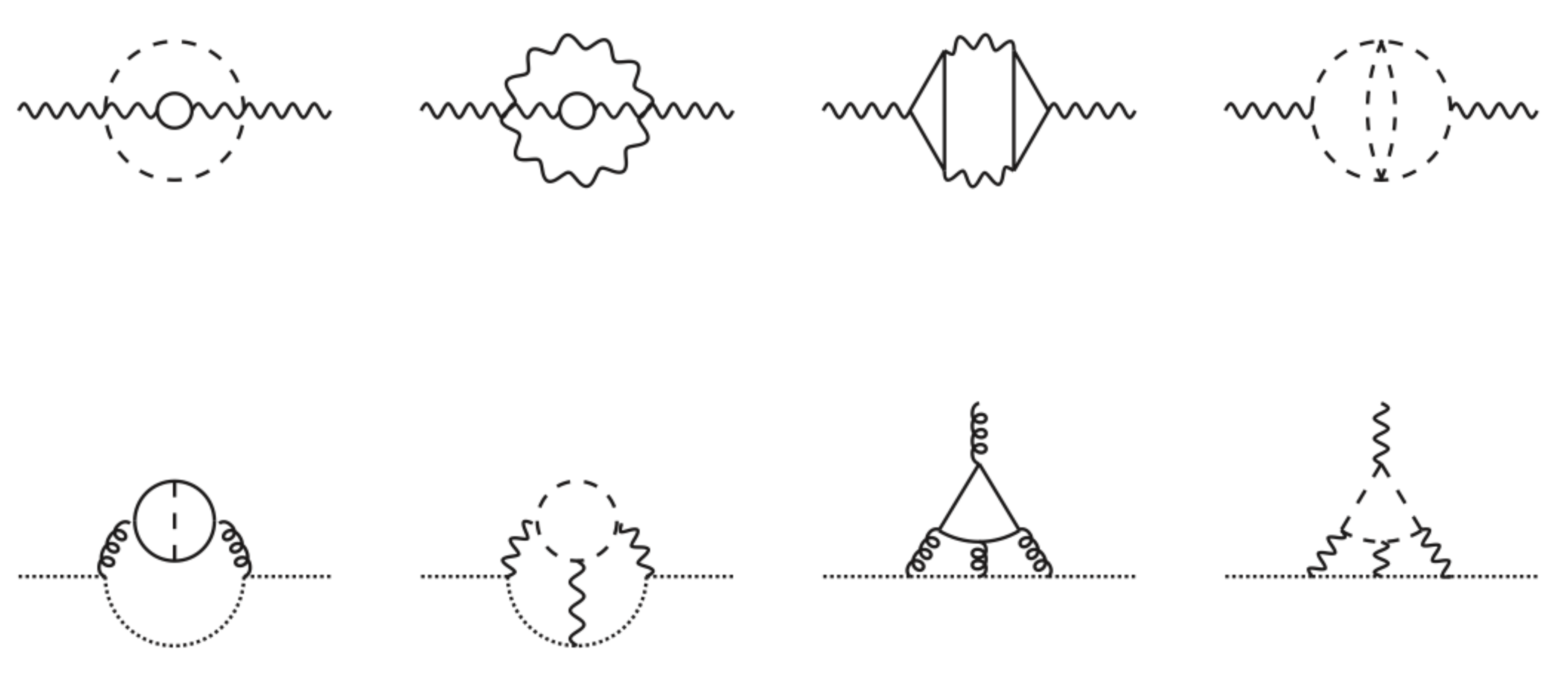}
\caption{\label{bf_dia} 
Sample of three-loop diagrams that appear in the calculation of $\beta$ functions.
Curly lines denote gauge bosons, dotted lines ghosts,
dashed lines scalar fields and solid lines fermions.}
\end{figure}

In Fig.~\ref{bf_dia} we show three-loop sample diagrams contributing to the considered two-
and three-point functions. For the explicit calculation of the required renormalization
constants, we use \msbar{} scheme  accompanied by
multiplicative renormalization. 
 As it has been shown in
Ref.~\cite{Chetyrkin:1984xa} the computation of the renormalization
constants in the \msbar{} scheme can be reduced to the evaluation of
only massless propagator  diagrams. The method
was successfully applied  to the
three-loop calculations of anomalous dimensions within \msbar{} or
\drbar{}
schemes~\cite{Larin:1993tp,Larin:1993tq,Pickering:2001aq,Harlander:2006rj,Harlander:2009mn,Chetyrkin:2012rz,Mihaila:2012fm}.              
 For the present calculation, we use a well-tested chain of programs: 
the Feynman rules of the model are obtained with the help of the program
 {\tt FeynRules}~\cite{Christensen:2008py} and translated into 
 {\tt QGRAF}~\cite{Nogueira:1991ex}  syntax. {\tt QGRAF} generates further all contributing Feynman
 diagrams. The 
output is passed via {\tt
  q2e}~\cite{Harlander:1997zb,Seidensticker:1999bb}, which transforms
Feynman diagrams into 
Feynman amplitudes, to {\tt exp}~\cite{Harlander:1997zb,Seidensticker:1999bb} 
that generates {\tt FORM}~\cite{Vermaseren:2000nd} code. The latter is
processed by {\tt   MINCER}~\cite{Larin:1991fz} 
that computes analytically massless propagator diagrams up to three
loops and outputs the $\epsilon$ expansion of the result.

The three-loop expressions for the beta functions of the gauge couplings in the
low-energy theory consisting in the SM and the electroweak triplets are
given through the following formulas:
\begin{widetext}
\begin{align}
\label{deltab1T}
 \beta_1^{\rm SM + T} &=\beta_1^{\rm SM}\nn \\
& +
\frac{\alpha _1^2}{\pi ^2} \left\{ \frac{\alpha _2^2 }{\pi ^2} C_{G_L}
  C_{R_L} N_{R_L} 
\left[  \left(   
   -\frac{11}{576} Y_f^2 N_f  -\frac{25}{576}  Y_h^2 N_h
 \right){N_{T_F}} 
%\right. \right.  \nn \\
%   & \left. 
+   
   \left( -\frac{23}{2304} Y_f^2 N_f -\frac{49}{2304} Y_h^2 N_h
   \right){N_{T_H}} \right]\right. \nn \\
   & \left. - \frac{\alpha _{\lambda_{hT}}^2}{\pi ^2} 
   \frac{1}{ 192}  Y_h^2 N_{R_L} N_h  N_{G_L} {N_{T_H}} 
 -\frac{\alpha _{\lambda_{hT}}}{\pi } \frac{\alpha _{\lambda_{h}}}{\pi }
  \frac{1}{8} Y_h^2 N_h^3+\frac{\alpha_2}{\pi } \frac{\alpha_{\lambda_{hT}}}{\pi }
 \frac{1}{24} C_{R_L} Y_h^2 N_h^2 + \frac{\alpha_1}{\pi }
 \frac{\alpha_{\lambda_{hT}}}{\pi }\frac{1}{8}  Y_h^4 N_h^2
\right\} \, ,
\end{align}
\end{widetext}
\begin{widetext}
\begin{align}
\label{deltab2T}
 \beta_2^{\rm SM + T} &= \beta_2^{\rm SM} +
\frac{\alpha _2^2}{\pi ^2} 
\left\{  C_{G_L}\left(
\frac{1}{6} {N_{T_F}} + \frac{1}{24}  {N_{T_H}} \right)
+ \frac{\alpha _2}{\pi} C_{G_L}^2 \left( \frac{1}{3} {N_{T_F}} + \frac{7}{48}  {N_{T_H}}
\right) \right. \nn \\
&  
+ \frac{\alpha _2^2}{\pi^2} 
\left[ 
\left( 
   \frac{247}{432} C_{G_L}^3
   -\frac{7}{108} C_{G_L}^2 T_{R_L} N_f
   -\frac{11}{576} C_{G_L} C_{R_L} T_{R_L} N_f
   \right. \right. \nn \\
   & \left.
   -\frac{127}{3456} C_{G_L}^2 T_{R_L} N_h
   -\frac{25}{576}  C_{G_L} C_{R_L} T_{R_L} N_h
   -\frac{145}{3456} C_{G_L}^3 {N_{T_F}}
   -\frac{277}{6912}C_{G_L} ^3 {N_{T_H}}   
\right)
{N_{T_F}} \nn \\
&  +
\left(
   \frac{2749}{6912} C_{G_L}^3
   -\frac{13}{432} C_{G_L}^2 T_{R_L} N_f
   -\frac{23}{2304} C_{G_L} C_{R_L} T_{R_L} N_f\right. \nn \\
   & \left. \left. \left. 
   -\frac{143}{6912} C_{G_L}^2 T_{R_L} N_h
   -\frac{49}{2304} C_{G_L} C_{R_L} T_{R_L} N_h  
   -\frac{145}{13824}  C_{G_L}^3 {N_{T_H}} \right) {N_{T_H}} \right] \right. \nn \\
& + \frac{\alpha _2}{\pi} \frac{\alpha _{\lambda_{T}}}{\pi} 
\frac{5}{64} C_{G_L}^2 {N_{T_H}^2}
+ \frac{\alpha _2}{\pi} \frac{\alpha _{\lambda_{hT}}}{\pi} \left(
\frac{1}{24 } C_{G_L}T_{R_L} N_h N_{T_H}  +  \frac{1}{32 } T_{R_L}^2 N_h^2 \right) 
- \frac{\alpha^2_{\lambda_{T}}}{\pi^2} \frac{1}{32}C_{G_L}(N_{G_L}+2) {N_{T_H}^3} \nn \\ 
& \left. 
+ \frac{\alpha^2_{\lambda_{hT}}}{\pi^2}  \left( -\frac{1}{96} C_{G_L} -\frac{1}{96} C_{R_L} \right) N_h {N_{T_H}} 
+ \frac{\alpha _1}{\pi} \frac{\alpha_{\lambda_{hT}}}{\pi} \frac{1}{48}
Y_h^2 T_{R_L} N_h^2 - \frac{\alpha_{\lambda_h} }{\pi} \frac{\alpha_{\lambda_{hT}}}{\pi} \frac{1}{16} T _{R_L} N_h^3
\right\} \, ,
\end{align}
\end{widetext}
%\begin{widetext}
\begin{multline}
\label{deltab3T}
\beta_3^{\rm SM + T} = \beta_3^{\rm SM} + 
\frac{\alpha _3^2}{\pi ^2} \frac{\alpha _2^2}{\pi ^2}C_{G_L} 
   C_{R_L}  N_{R_L}T_{R_C} N_q \\ 
   \times \left(-\frac{11}{576}  {N_{T_F}}
   -\frac{23}{2304} {N_{T_H}}\right) \, .
\end{multline}
%\end{widetext}
In the above equations $\beta_i^{\rm SM}$ denote the beta functions of
the gauge couplings in the SM that can be found in
Refs.~\cite{Mihaila:2012fm,Mihaila:2012pz}. Furthermore,  
we use the following abbreviations: 
$Y_f^2 N_f =  N_{R_C} Y_q^2 N_q  + Y_\ell^2 N_\ell $  and  $N_f=
N_{R_C} N_q + N_\ell $.
The numerical values of the beta functions 
specified to our case are obtained by means of the following replacements: 
(i) $Y_q = \sqrt{\tfrac{3}{5}}\tfrac{1}{6}$, 
$Y_\ell = -\sqrt{\tfrac{3}{5}}\tfrac{1}{2}$, 
$Y_h = \sqrt{\tfrac{3}{5}}\tfrac{1}{2}$ denoting the hypercharges of
the SM quarks, leptons and Higgs in the SU(5) normalization;
(ii) $N_q = N_\ell = 3$, $N_h = 1$
and $N_{T_{F,H}}=1$ standing for the number of SM quark and lepton generations, Higgs
and electroweak triplets. To recover the  expressions  for the beta
functions in the notation of
Refs.~\cite{Mihaila:2012fm,Mihaila:2012pz}, we have to make the
replacement $N_q = N_\ell = n_g$.

In order to cross-check our results, we reproduced with our setup the
results for the three-loop gauge beta functions of the SM. Let us 
mention that we use a different implementation than the one of
Refs.~\cite{Mihaila:2012fm,Mihaila:2012pz} based on complete multiplets wrt the SM gauge group,
{\it e.g.}~left-handed leptons populating 
the ${\rm SU(2)}_L$ doublet $\ell$ are treated as the same particle in the loops, 
thus exploiting the full ${\rm SU(2)}_L$ symmetry of the unbroken SM phase.  
When available, we also compared the contributions generated by the electroweak
triplets in the gauge sector with the results  of
Ref.~\cite{Pickering:2001aq} and obtained complete agreement. 

Furthermore, the contributions of the colour octets to the beta functions 
in the SM+T+O model can be
read from \eqs{deltab1T}{deltab3T}  
after the proper substitutions. We give the results
explicitly in \app{Ocontbf}.

%===========================================================
\subsection{Decoupling coefficients}
\label{deccoeff}
%===========================================================
In this section we describe the calculation of the two-loop 
decoupling coefficients for the SM gauge couplings when the electroweak triplets 
$T_{H,F}$ are integrated out. 
We present our results again in terms of group-theory invariants,  
so that our calculation can be generalized to other gauge groups as well.

Let us define at this point the decoupling coefficients for the gauge
couplings when the SM+T model is matched with the SM,
\be
\label{defdc}
\alpha'_i (\mu) = \zeta_{\alpha_i} \left( \mu, \alpha_i(\mu), m_{T_{F,H}} (\mu)  \right) \alpha_i (\mu)\,.
\ee
Here $\mu$ denotes the scale at which the decoupling of the electroweak
triplets is performed. It is not fixed by the theory, but it is usually
chosen of the order of the electroweak-triplet masses. It is expected
that the dependence of the physical observables on this unphysical
parameter is reduced order by order in perturbation theory. Such an
example is illustrated in~\fig{FIGalpha2atMG}, in the next section.  The
parameters on the right-hand side of the equality are all defined in the
SM+T model, whereas the SM parameters are labeled with a
prime.

\begin{figure}[ht]
\includegraphics[angle=0,width=8.5cm]{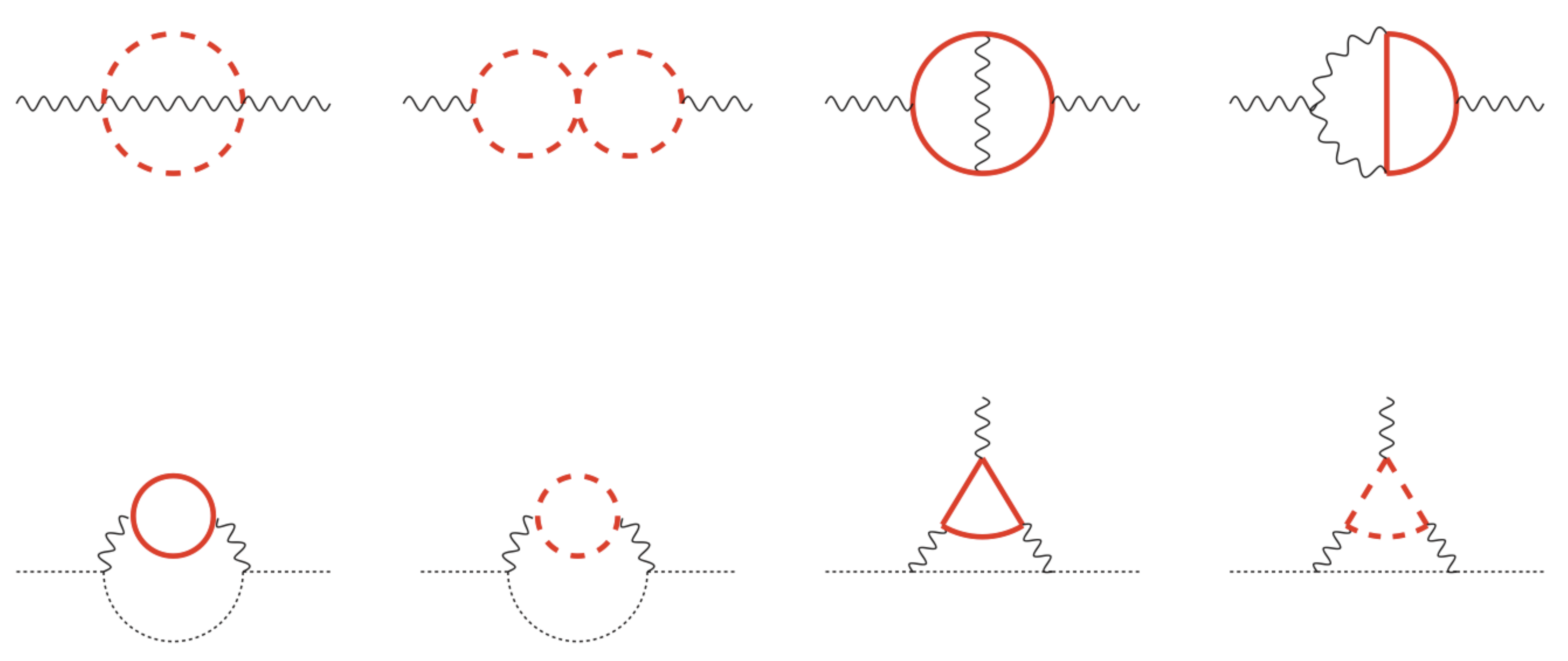}
\caption{\label{dec_dia} 
Sample two-loop diagrams that appear in the calculation of $\zeta_{\alpha_2}$.
      Red (bold) lines represent massive (scalar and fermionic) triplets  
      and black (thin) lines massless fields.
      Furthermore, curly lines denote gauge bosons, dotted lines ghosts, dashed lines scalar fields and
      solid lines fermions.}
\end{figure}

For the computation of the coefficient $\zeta_{\alpha_i}$ one has to
consider Green's functions involving light particles and a vertex that contains
the gauge coupling $\alpha_i$. Since the matching coefficients are
universal quantities, they must be independent of the momentum transfer
of the specific process taken under consideration. Ref.~\cite{Chetyrkin:1997un} showed
that the matching coefficients for the gauge couplings can be calculated
from the gauge bosons and Fadeev-Popov ghost propagators and from the gauge
boson-ghost vertex, all evaluated at vanishing external momenta. 
Thus, in dimensional regularization only diagrams containing at least
one heavy particle inside the loops contribute and only the hard
regions in the asymptotic expansion of the diagrams have to be taken
into account. We show in Fig.~\ref{dec_dia} sample two-loop Feynman diagrams
contributing to the matching coefficient for the gauge coupling
$\alpha_2$. The Feynman diagrams are computed within our setup with the
same chain of automated programs as for the calculation of the beta
functions, except for the fact that the resulting   Feynman amplitudes
are mapped to two-loop massive tadpole topologies that are handled with
the help of the program MATAD~\cite{Steinhauser:2000ry}.

When the electroweak triplets are integrated out, only the gauge
coupling $\alpha_2$ is modified. Its decoupling coefficient up to two
loops reads
\begin{widetext}
\begin{align}
\label{zetaalpha2}
\zeta_{\alpha_2} &=
1 + \frac{\alpha_2}{\pi} C_{G_L} \left( -\frac{1}{6}  \ln{\frac{\mu^2}{m_{T_F}^2}} {N_{T_F}} 
-\frac{1}{24}  \ln{\frac{\mu^2}{m_{T_H}^2}} {N_{T_H}} \right) \nn \\
& + \left(\frac{\alpha_2}{\pi}\right)^2 C_{G_L}^2 \left[ 
\left( 
-\frac{7}{288} 
-\frac{1}{12}  \ln{\frac{\mu^2}{m_{T_F}^2}}
+\frac{1}{36}  \ln^2{\frac{\mu^2}{m_{T_F}^2}} {N_{T_F}}
+ \frac{1}{72}  \ln{\frac{\mu^2}{m_{T_F}^2}} \ln{\frac{\mu^2}{m_{T_H}^2}} {N_{T_H}} 
\right) {N_{T_F}} \right. \nn \\
& \left. 
+ \left(\frac{37}{576} 
-\frac{11}{96}  \ln{\frac{\mu^2}{m_{T_H}^2}} 
+\frac{1}{576}  \ln^2{\frac{\mu^2}{m_{T_H}^2}} {N_{T_H}} \right)
{N_{T_H}} \right]
% \nn \\ & 
+ \frac{\alpha _2}{\pi}\frac{\alpha _{\lambda _T}}{\pi}C_{G_L}(N_{G_L}+2)
\left(
-\frac{1}{48} 
-\frac{1}{48}  \ln{\frac{\mu^2}{m_{T_H}^2}} 
\right) {N_{T_H}^2} \, ,
\end{align}
\end{widetext}
where the  masses $m_{T_{F,H}}$ in \eq{zetaalpha2} are defined in the $\overline{\text{MS}}$ 
scheme. The anomalous dimensions $\gamma_{m_{T_{F,H}}}$, 
governing their scale dependence, are defined through
\be
\label{runmass}
\mu^2 \frac{d}{d \mu^2} m_{T_{F,H}} = m_{T_{F,H}} \gamma_{m_{T_{F,H}}} \, .
\ee 
At one-loop order, they read
\begin{align}
\label{gammaTF}
\gamma^{(1-\rm{loop})}_{m_{T_F}} &= -\frac{\alpha_2}{\pi} \frac{3}{4} C_{G_L} \, , \\
\label{gammaTH}
\gamma^{(1-\rm{loop})}_{m_{T_H}} &= -\frac{\alpha_2}{\pi} \frac{3}{8} C_{G_L} +
\frac{\alpha_{\lambda_{T}}}{\pi} \frac{N_{G_L}+2}{2} \, .
\end{align}
The numerical value of the decoupling coefficients 
specified to our case is obtained by means of the group invariants given
in \Table{tab:inv} and by setting $N_{T_{F,H}} = 1$. The one-loop contributions agree with the well-known
result computed for the first time in Ref.~\cite{Hall:1980kf}. 
The two-loop results given in \eq{zetaalpha2} are new.

The contribution of the colour octets to $\zeta_{\alpha_3}$ can be 
derived in a similar manner. It can also be
read from \eq{zetaalpha2}, 
after the proper substitutions (cf.~\app{Ocontmatch}) for the group invariants.  

%%%%%%%%%%%%%%%%%%%%%%%%%%%%%%%%%%%%%%%%%%%
\section{Numerical analysis}
\label{numan}
%%%%%%%%%%%%%%%%%%%%%%%%%%%%%%%%%%%%%%%%%%%

%%%%%%%%%%%%%%%%%%%%%%%%%%%%%%%%%%%%%%%%%%%%%%%%%%%%%%%%%%%%%%
In this section we study the numerical impact of the newly computed corrections 
on the evolution of the gauge couplings and on the correlation function
between the electroweak-triplet masses and the GUT scale. In practice,
 we integrate numerically the $n$-loop beta functions of the 
gauge couplings taking into account also the $(n-1)$-loop 
running of the top-Yukawa 
coupling and the $(n-2)$-loop running of the Higgs boson self-coupling.
We can safely neglect the contribution of the bottom and tau Yukawa couplings  
and defer the study of the effect due to the new scalar self-interactions 
of the scalar triplet $T_H$ to \sect{scalarsi} (i.e.~we set here
$\alpha_{\lambda_T} = 0$ and $\alpha_{\lambda_{hT}} = 0$).  
As input parameters for the running analysis we take \cite{Mihaila:2012pz}
\begin{align}
\label{alpha1MZ}
\alpha_1^{\overline{\text{MS}}}  (M_Z) &= 0.0169225 \pm 0.0000039 \, ,  \\
\label{alpha2MZ}
\alpha_2^{\overline{\text{MS}}}  (M_Z) &= 0.033735 \pm 0.000020 \, ,  \\
\label{alpha3MZ}
\alpha_3^{\overline{\text{MS}}}  (M_Z) &= 0.1173 \pm 0.00069 \, ,  \\
\label{alphatMZ}
\alpha_t^{\overline{\text{MS}}}  (M_Z) &= 0.07514 \, ,
\end{align}
given in the full SM, i.e.~with the top quark threshold effects taken into account.\footnote{See \cite{Martens:2010nm} for a description of how these quantities are obtained from their 
experimental counterparts \cite{Beringer:1900zz}.} 
 The Higgs self-coupling in \eq{Vren} is determined assuming a Higgs
 boson with mass $125$ GeV. Thus, we obtain
\be
\label{allamH}
\alpha_{\lambda_h} \approx 0.010 \, . 
\ee
Let us start by studying the impact of the electroweak triplets on the 
running and decoupling of $\alpha_2$. When the decoupling is performed at the two-loop level 
the running masses must be consistently evolved at the one-loop order. 
At that order \eq{runmass} can be easily integrated analytically yielding 
\be
m_{T_{F,H}} (\mu) = m_{T_{F,H}} (\mu_0) \left( \frac{\alpha_2 (\mu)}{\alpha_2 (\mu_0)} 
\right)^{- \frac{\pi}{\alpha_2 a_2} \gamma^{(1-\rm{loop})}_{m_{T_{F,H}}} } \, ,
\ee
where $a_2$ is the one-loop coefficient of the ${\rm SU(2)}_L$ gauge coupling beta function 
defined in \eq{defbf}, whereas $\gamma^{(1-\rm{loop})}_{m_{T_{F,H}}}$ are given in \eqs{gammaTF}{gammaTH}.
In the following we will drop for simplicity the scale dependence of the running masses. 
Unless otherwise specified the symbol ``$m$'' should be understood as $m (\mu = m)$.

\begin{figure}[ht]
\includegraphics[width=8.cm]{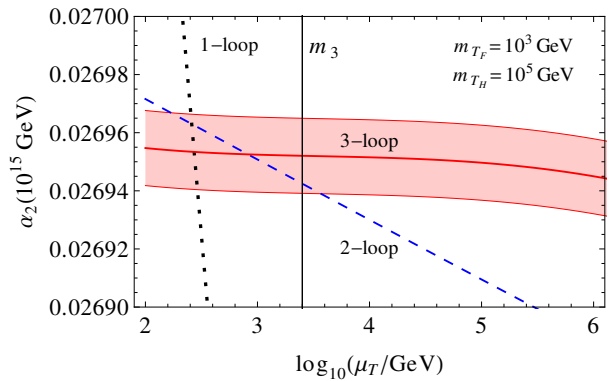}
\caption{\label{FIGalpha2atMG}
The value of the coupling $\alpha_2$ evaluated at the reference scale 
of $10^{15} \, \text{GeV}$ is shown as a function of the decoupling scale 
of the triplets $\mu_T$. 
Dotted (black), dashed (blue) and full (red) lines correspond respectively to the one-, two- and three-loop 
running analysis. The $1\sigma$ error band coming from the experimental value 
of $\alpha_2 (M_Z)$ is shown for the three-loop curve as well. 
The vertical (black) line denotes the quantity $m_3$ (cf.~\eq{defm3}), 
corresponding to the ``naive'' decoupling scale chosen in the two-loop analysis.  
}
\end{figure}

In \fig{FIGalpha2atMG}, we plot the gauge coupling $\alpha_2$ evolved until 
the reference scale of $10^{15}$ GeV as a function of the (unphysical) 
decoupling scale $\mu_T$ where the electroweak triplets are integrated out.
We expect that the dependence on this unphysical parameter is reduced
order by order in perturbation theory. Thus, we can use it as a measure
of the convergence of the perturbation expansion. Indeed, 
from \fig{FIGalpha2atMG} we observe that the scale dependence is drastically reduced when the 
three-loop corrections are taken into account. Let us mention that the prediction  
obtained from the two-loop analysis for the ``natural'' choice of the decoupling 
scale $\mu_T = m_3$,\footnote{One can verify 
(cf.~\eq{zetaalpha2}) that for $\mu_T = m_3$ the one-loop contribution 
to $\zeta_{\alpha_2}$ vanishes.} where 
\begin{equation}
\label{defm3}
m_3 \equiv \left(m_{T_F}^4 m_{T_H}\right)^{1/5} \, ,  
\end{equation}
is usually within the experimental band of the three-loop result. For this choice of scale, 
discrepancies between the two- and three-loop predictions beyond the experimental accuracy 
are obtained only in the hierarchical case $m_{T_H} \ll m_{T_F}$.\\
Analogous considerations hold also for the effects of the colour-octet
states $O_{F,H}$ on the running and decoupling of $\alpha_3$.
However, the larger experimental uncertainty on $\alpha_3 (M_Z)$ 
always dominates over the theoretical mismatch between the two- and three-loop predictions.

\begin{figure}[ht]
\includegraphics[width=8.cm]{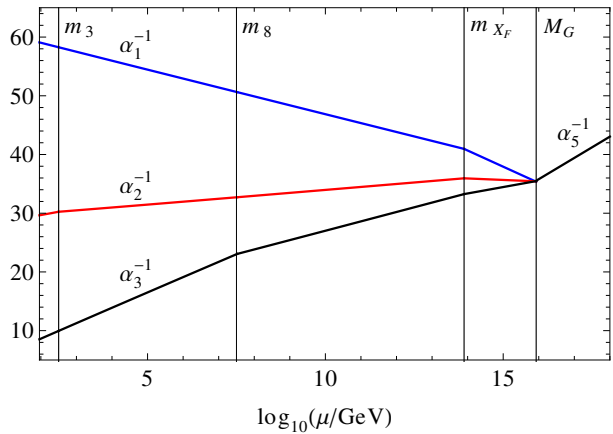}
\caption{\label{sampleunif}
Sample three-loop unification pattern for 
$m_{T_F} = m_{T_H} = 10^{2.5} \, \text{GeV}$, 
$m_{O_F} = m_{O_H} = 10^{7.5} \, \text{GeV}$,
$m_{X_F} = M_G / 100$ and 
$m_{\mathcal{T}} = m_{X_V} = M_G$. 
The lines with different slopes from top to bottom correspond to 
$\alpha_1^{-1}$ (blue), $\alpha_2^{-1}$ (red) and $\alpha_3^{-1}$ (black). 
The dashed vertical lines denote the masses of the intermediate-scale thresholds, 
where $m_3$ is defined in \eq{defm3} and $m_8$ is analogously defined as 
$m_8 \equiv \left(m_{O_F}^4 m_{O_H}\right)^{1/5}$.
}
\end{figure}

In \fig{sampleunif} we show a sample three-loop unification pattern for
the inverse of the gauge couplings and for the following choice of the intermediate-scale thresholds: 
$m_{T_F} = m_{T_H} = 10^{2.5} \, \text{GeV}$, 
$m_{O_F} = m_{O_H} = 10^{7.5} \, \text{GeV}$,
$m_{X_F} = M_G / 100$ and 
$m_{\mathcal{T}} = m_{X_V} = M_G$. 
Here $M_G$ is operatively defined as the scale at which $\alpha_1$, $\alpha_2$ and $\alpha_3$ meet, 
up to GUT-threshold corrections. 
The running and decoupling procedure is performed at the NNLO level
(i.e.~three-loop running and two-loop matching)  
with the exception of the short final stage of the running between
$m_{X_F}$ and $M_G$ for which we consider the  
decoupling of $X_F$ and its contribution to the gauge coupling beta functions only at 
the one- \cite{Weinberg:1980wa,Hall:1980kf} and two-loop level \cite{Machacek:1983tz}, respectively. 
Furthermore, the GUT-threshold corrections are considered only at one loop \cite{Weinberg:1980wa,Hall:1980kf}.

In order to quantify the impact of the newly computed corrections 
let us mention that for such a sample unification pattern 
the relative difference between the two- and three-loop values 
of $\alpha_1$, $\alpha_2$ and $\alpha_3$ evaluated at $M_G$ 
amounts to $0.015 \%$, $0.061 \%$ and $0.08 \%$, respectively. 
This has to be compared with the  relative experimental uncertainties:
$\Delta \alpha_1 / \alpha_1 = 0.023 \%$, $\Delta \alpha_2 / \alpha_2 = 0.059 \%$ 
and $\Delta \alpha_3 / \alpha_3 = 0.59 \%$. 
Hence, for $\alpha_1$ and $\alpha_2$ the three-loop corrections are 
of the same order of magnitude as the experimental uncertainties, 
while for the case of $\alpha_3$ the experimental error  dominates 
with respect to the theoretical one. 

\begin{figure}[ht]
\includegraphics[width=8.cm]{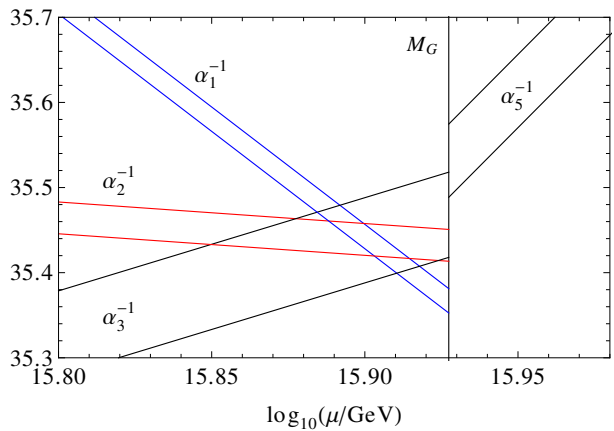}
\caption{\label{sampleunif_zoom}
Detail of the three-loop unification pattern of \fig{sampleunif} in the vicinity of the unification scale, 
including the $1\sigma$ error bands. 
}
\end{figure}

In~\fig{sampleunif_zoom}, the region of gauge coupling unification is enlarged.
The inverse of the coupling constants $\alpha_1, \alpha_2$ and $\alpha_3$ is shown together with the
error bands induced by the experimental uncertainties on 
their values at the scale $M_Z$. 
From the figure it is evident that threshold effects at the GUT scale 
have to be taken into account for a proper unification. 

One of the most interesting observables of the present running analysis is  
the correlation between the effective electroweak-triplet mass $m_3$ (cf.~\eq{defm3}) 
and the unification scale $M_G$ \cite{Bajc:2006ia}. Such a correlation mainly depends on the convergence scale 
of the couplings 
$\alpha_1$ and $\alpha_2$, whenever the masses of the super-heavy particles $X_F$ and $\mathcal{T}$ are fixed. 
The coupling $\alpha_3$ enters the correlation function only indirectly from the two-loop level on. 
Hence, for this particular observable, the uncertainty induced by $\alpha_3 (M_Z)$ remains always subleading  
wrt that caused by $\alpha_{1,2} (M_Z)$. Moreover, also the colour-octet states $O_{F,H}$, 
that give sizable contributions only to the evolution of the strong coupling constant, have a minor role. \\
The predicted value for the couplings $\alpha_{1,2}$ at high energies 
maintains an exact dependence on $m_3$ at the one- and two-loop order (whenever the triplets are decoupled at the scale 
$\mu_T = m_3$) and remains approximate, 
usually within the experimental uncertainty, at the three-loop level. 

The upper bound on the effective triplet mass $m_3$ 
is the crucial parameter for phenomenology \cite{Bajc:2006ia}. 
For a fixed unification scale, $m_3^{\rm{max}}$ is obtained by maximizing the masses of the extra thresholds 
$X_F$ and $\mathcal{T}$. 
For a given choice of the cutoff $\Lambda$ of the SU(5) effective theory, namely $\Lambda = 100 \ M_G$ 
(cf.~the discussion at the end of \sect{let}), the masses of the electroweak triplets are 
maximized by taking $m_{X_F} = M_G / 100$. For the colour-triplet
scalar $\mathcal{T}$ the maximal allowed mass scale is in principle the
Planck scale. However, the dependence of $m_3^{\rm{max}}$ on the colour-triplet 
mass turns out to be mild. For instance, varying the colour-triplet mass between the
unification and the Planck scales induces a variation on the parameter
$m_3^{\rm{max}}$ which lays within the experimental uncertainty. 
For convenience, we set the mass of the colour-triplet scalar to the
unification scale $m_{\mathcal{T}} = M_G$. 
Here, $M_G$ is operatively defined as the scale where $\alpha_1$ and $\alpha_2$ meet 
up to corrections induced by the one-loop matching between the 
$\rm{SM+T+O+X_F}$ and the ${\rm SU(5)}+24_F$ theories. 
The size of these matching corrections can be read from~\fig{sampleunif_zoom}. 
We also identify $M_G$ with the mass of the super-heavy gauge boson $X_V$ 
responsible for the gauge-induced proton decay rate.

\begin{figure}[ht]
\includegraphics[width=8.cm]{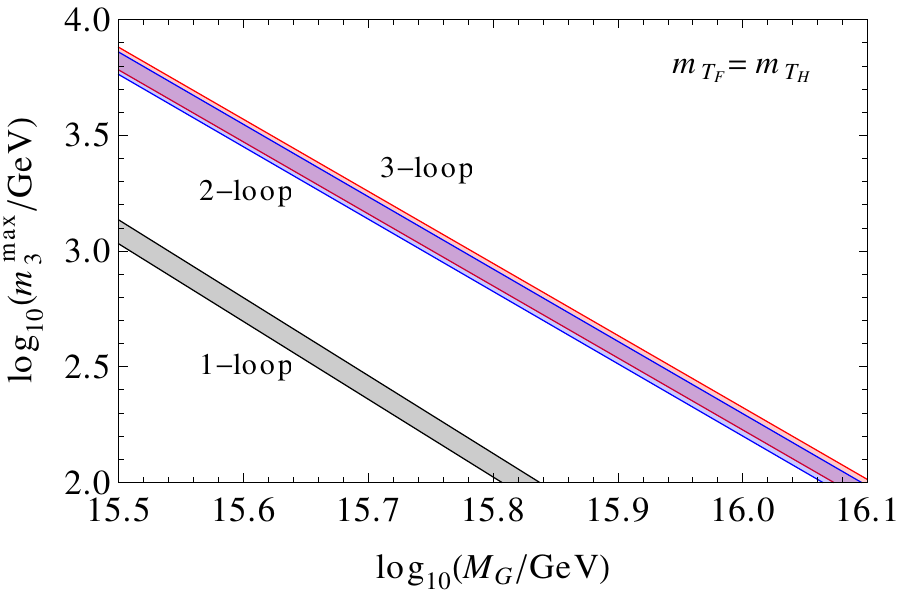}
\caption{\label{MGvsm3} 
The maximal value of the effective triplet mass $m_3$ as a function of the unification scale $M_G$. 
The black, blue and red bands (from bottom-left to top-right) correspond respectively to the one-, two- and three-loop 
running analysis. The three-loop result is obtained by taking $m_{T_F} = m_{T_H}$. 
The error bands are due to the $1\sigma$ uncertainties on the low-energy couplings $\alpha_1 (M_Z)$  
and $\alpha_2 (M_Z)$ (cf.~\eqs{alpha1MZ}{alpha2MZ}).
%$1\sigma$ error bands.
}
\end{figure}

In \fig{MGvsm3} we show $m_3^{\rm{max}}$ as a function of  $M_G$  at the 
one-, two- and three-loop level, respectively. 
Notice that the two-loop correction on $m_3^{\rm{max}}$ for a fixed $M_G$ 
is of the same order of magnitude as the one-loop contribution and amounts to several TeV. 
On the other hand, the three-loop correction pushes the correlation only
a bit up, 
but always within the experimental uncertainty of the two-loop band. 
Hence, the theoretical error 
due to the perturbative expansion (defined by the relative 
difference between the $n$- and the $(n-1)$-loop prediction)
is reduced now at the same level as the experimental uncertainty induced by the 
measurement of $\alpha_1$ and $\alpha_2$ at the Z-boson mass scale. 
From \fig{MGvsm3} we can estimate that 
for a given unification scale $M_G$, the effective parameter 
$m_3^{\rm{max}}$ can be now determined with a $25 \%$ accuracy.  

\begin{figure}[ht]
\includegraphics[width=8.cm]{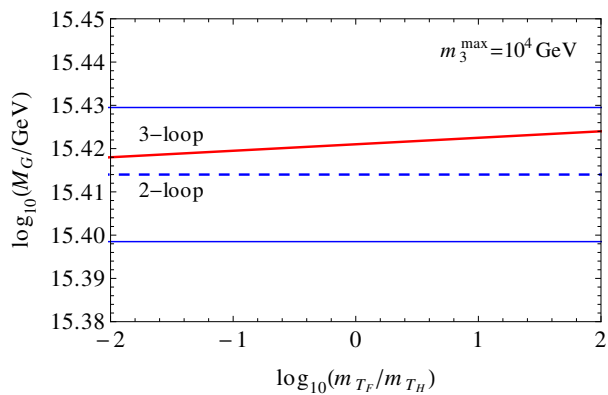}
\caption{\label{MGvsFHratio} 
$M_G$ as a function of the $m_{T_F} / m_{T_H}$ ratio 
for a fixed value of $m_3^{\rm{max}}$.
The dashed (blue) and full (red) lines 
correspond respectively to the two- and three-loop 
running analysis. The $1\sigma$ error band is shown as well for the two-loop case. 
The negative and positive extrema on the x-axis correspond respectively to the configurations 
$m_{T_F} = 10^{3.6} \ \text{GeV}$, $m_{T_H} = 10^{5.6} \ \text{GeV}$ and 
$m_{T_F} = 10^{4.4} \ \text{GeV}$, $m_{T_H} = 10^{2.4} \ \text{GeV}$. 
}
\end{figure}

It is worth mentioning that starting from three loops 
the $m^{\rm{max}}_3 - M_G$ correlation shows also a dependence 
on the ratio $m_{T_F} / m_{T_H}$. 
In particular, larger deviations between the two- and three-loop analysis are observed 
for the case when $m_{T_H} \ll m_{T_F}$. However, 
for the  mass range relevant for unification, the three-loop corrections
to   the correlation function $m^{\rm{max}}_3 - M_G$
for $m_{T_H} \neq m_{T_F}$ turn out to be always within the experimental 
uncertainty of the two-loop prediction.
 For illustration, we show in \fig{MGvsFHratio}  $M_G$ as a function of the $m_{T_F} / m_{T_H}$ ratio 
for a fixed value of $m_3^{\rm{max}}$.

Finally, the analysis of the full unification pattern, including the convergence of 
$\alpha_3$ with $\alpha_{1,2}$, fixes the masses of the colour octets $O_{F,H}$ 
in terms of the other masses of the model (see the discussion in Ref.~\cite{Bajc:2006ia,Bajc:2007zf}). 
However, the newly computed corrections induced by the colour-octet states to the two-loop matching 
coefficient and three-loop beta function of $\alpha_3$ (cf.~\app{extrares}) 
are subleading as compared to the experimental uncertainty on $\alpha_3 (M_Z)$.  
A two-loop analysis as in \cite{Bajc:2006ia,Bajc:2007zf} is usually sufficient.

%===========================================================
\subsection{Scalar self-interactions}
\label{scalarsi}
%===========================================================

At the tree-loop level all the sectors of the theory enter for the first time 
into the running of the gauge couplings. In particular, this is also true  
for the couplings $\alpha_{\lambda_h}$, $\alpha_{\lambda_T}$ and $\alpha_{\lambda_{hT}}$ 
of the scalar potential in \eq{Vren}. 
However, while $\alpha_{\lambda_h}$ is fixed in terms of the SM Higgs boson mass 
(cf.~\eq{allamH}), $\alpha_{\lambda_T}$ and $\alpha_{\lambda_{hT}}$ are essentially 
unconstrained\footnote{Notice that $\alpha_{\lambda_{hT}}$ does modify 
the decay properties of the Higgs boson (see e.g.~\cite{Chang:2012ta}). 
However, one would expect that such an effect can be 
arbitrarily suppressed for large enough $m_{T_H}$.} 
and could, if large enough, contribute significantly to the running of 
the gauge couplings after $T_H$ is integrated in. 

In order to quantify how large the scalar self-couplings can be, 
let us inspect their one-loop beta functions \cite{Forshaw:2003kh} 
\begin{align}
\label{betah}
\pi ^2 \beta_{\alpha_{\lambda_{h}}} &= 
 3 \alpha _{\lambda _h}^2+\frac{3}{2} \alpha _{\lambda _h} \alpha _t
+\frac{3}{4} \alpha_{\lambda _{hT}}^2-\frac{3}{4} \alpha _t^2 
\, , \\
\label{betaT}
\pi ^2 \beta_{\alpha_{\lambda_{T}}} &= 
\frac{11}{2} \alpha _{\lambda_T}^2
+ \frac{1}{2}\alpha _{\lambda _{hT}}^2
\, , \\
\label{betahT}
\pi ^2 \beta_{\alpha_{\lambda_{hT}}} &= 
\alpha _{\lambda _{hT}}^2 + \left( \frac{3}{2} \alpha _{\lambda _h} 
+\frac{5}{2} \alpha _{\lambda _T} +\frac{3}{4} \alpha _t\right) \alpha _{\lambda _{hT}}
\, ,
\end{align}
where we considered for simplicity the gauge-less limit $(\alpha_{1,2} \rightarrow 0)$ and 
retained only the top-Yukawa contribution $\alpha_t$. 
The definition of the beta functions follows the conventions in \eq{defbf}.   

\eqs{betah}{betahT} show that for large and positive initial values of the couplings 
$\alpha_{\lambda_T}$ and $\alpha_{\lambda_{hT}}$ 
the renormalization group evolution is such that all the scalar 
self-couplings can easily become nonperturbative below 
$M_G$.
In such a situation perturbation theory breaks down,\footnote{It is in principle conceivable that the 
inclusion of extra interactions which are remnants of the complete GUT theory 
(cf.~\app{lowenint}) could stabilize the scalar potential and bring the 
couplings back to the perturbative regime. The study of such a scenario, however, is beyond 
the scopes of our work.} 
meaning that we cannot trust our predictions about gauge coupling unification.  
Imposing the conservative bound $\alpha_{\lambda_T,\lambda_{hT}} < 0.01$ we have checked that 
Landau poles are not developed below $M_G$ and the effects on the $m_3^{\rm{max}} - M_G$ 
correlation are always within the experimental band of the two-loop analysis.

%%%%%%%%%%%%%%%%%%%%%%%%%%%%%%%%%%%%%%%%%%%
\section{Conclusions and outlook}
\label{concl}
%%%%%%%%%%%%%%%%%%%%%%%%%%%%%%%%%%%%%%%%%%%

In this work we have undertaken an important step towards 
the study of gauge coupling unification in the ${\rm SU(5)}+24_F$ model~\cite{Bajc:2006ia,Bajc:2007zf} 
at the three-loop level. 
We computed the contributions 
of the electroweak triplets $T_{F,H}$ and the colour octets $O_{F,H}$ 
(which are predicted to be well below the GUT scale in this specific model) 
to the three-loop beta functions and the two-loop matching coefficients
of the SM gauge couplings. 

In particular, 
the most important observable of the running analysis is the correlation between 
the maximal value of an effective triplet mass parameter $m^{\rm{max}}_3$ and the 
unification scale $M_G$. This correlation is shown in \fig{MGvsm3} and is such that the 
electroweak triplets can escape the detection at LHC only if the unification scale is below $\approx 10^{16}$ GeV, 
thus implying a proton lifetime which should be accessible to the future generation of megaton-scale 
proton decay experiments \cite{Abe:2011ts}.\\ 
Such a correlation needs to be computed as accurately as possible. 
Indeed, for a fixed value of $M_G$, the parameter $m^{\rm{max}}_3$ can be in principle extracted 
from the low-energy values of $\alpha_1$ and $\alpha_2$ with an accuracy
of about $25 \%$. 
On the other hand, for a fixed value of $M_G$, the values of
$m^{\rm{max}}_3$ predicted at one and two loops differ by around $100 \%$ (cf.~again \fig{MGvsm3}). 
Hence, for this particular observable which is almost insensitive 
to the low-energy value of $\alpha_3$, the three-loop corrections are
required in order to settle the accuracy of the theoretical prediction 
at the level of the experimental precision. 

Still, one should keep in mind the existence of irreducible theoretical 
uncertainties which plague any GUT and may endanger the predictivity of a given scheme. 
These are, for instance, the presence of effective operators which are required either by the 
self-consistency of the theory (as in the ${\rm SU(5)}+24_F$ model) 
or which are expected on physical grounds due to the vicinity 
of the Planck and the GUT scales.\footnote{For an SO(10) model where such effects are 
less relevant see \cite{Bertolini:2013vta,*Bertolini:2012im,*Bertolini:2009es}.} 
In this sense, an effort towards a three-loop analysis of gauge coupling unification 
should be minimally understood as a way to reduce the theoretical error due to the 
perturbative expansion. 

However, the path towards a complete three-loop analysis of gauge coupling unification 
in GUTs (with or without supersymmetry) is still long and many important ingredients are still missing. 
These are, for instance, the contributions to the three-loop beta functions and 
the two-loop matching coefficients of arbitrary multiplets charged under the SM group. 
Intermediate-mass scale multiplets are usually predicted in nonsupersymmetric GUTs 
and the knowledge of a general formula for their contribution 
could allow to extend the study of gauge coupling unification at the three-loop level 
also to other well motivated scenarios based on 
SU(5) \cite{Dorsner:2005fq,*Dorsner:2005ii,Perez:2007rm,Dorsner:2007fy,Feldmann:2010yp} 
and SO(10) \cite{Bertolini:2013vta,*Bertolini:2012im,*Bertolini:2009es}. 
Finally, the last important and conceptually challenging ingredient 
is represented by the two-loop matching at the GUT scale. 
In this respect, a step towards such a calculation has already been 
performed in the context of the GG SU(5) model \cite{Martens:2010pe} 
and could be in principle extended both to nonsupersymmetric and supersymmetric GUTs.

%%%%%%%%%%%%%%%%%%%%%%%%%%%%%%%%%%%%%%%%%%%
\subsection*{Acknowledgments}
%%%%%%%%%%%%%%%%%%%%%%%%%%%%%%%%%%%%%%%%%%%

We thank Borut Bajc, Miha Nemev\v{s}ek and Goran Senjanovi\'{c} 
for their interest in this project and for useful discussions. 
This work was supported by the DFG through the SFB/TR 
9 ``Computational Particle Physics''. 

\appendix

%%%%%%%%%%%%%%%%%%%%%%%%%%%%%%%%%%%%%%%%%%%
\section{Details of the SU(5) + ${\bf 24_F}$ model}
\label{details24F}
%%%%%%%%%%%%%%%%%%%%%%%%%%%%%%%%%%%%%%%%%%%

In this appendix we collect some basic facts about the SU(5) model augmented with a 
fermionic $24_F$ multiplet. In particular, we recompute the mass spectrum and 
derive the low-energy interactions among 
the SM fields and the remnant GUT states which populate the desert at intermediate mass scales.  
The latter motivate the interactions included into the three-loop analysis computation.  

%===========================================================
\subsection{Field content and SM embedding}
\label{FCandSMem}
%===========================================================

The field content of the model features an additional $24_F$ 
on top of the original representations of the GG model, namely 
three copies of $\overline{5}_F \oplus 10_F$ and $5_H \oplus 24_H$ 
in the Higgs sector. The embedding of the SM fields into the SU(5) 
representations is symbolically displayed in \sect{SU524Fmodel}. 
More precisely, spanning the SU(5), ${\rm SU(3)}_C$ and ${\rm SU(2)}_L$ 
spaces respectively with latin ($a = 1, \ldots , 5$), greek ($\alpha = 1, \ldots , 3$) 
and capital-latin ($A = 1, 2$) letters, we have 
\be
(5_{H})_a = 
\left(
\begin{array}{c}
\mathcal{T}_\alpha \\ 
h_A 
\end{array}
\right) \, ,
%\ee
\qquad
%\be
(\overline{5}_F)^a = 
\left(
\begin{array}{c}
d^{c\alpha} \\ 
\epsilon^{AB} \ell_B 
\end{array}
\right) \, ,
\ee
\be
(\overline{10}_{F})_{ab} = \tfrac{1}{\sqrt{2}}
\left(
\begin{array}{cc}
\epsilon_{\alpha\beta\gamma} u^{c\gamma} & - q_{\alpha B} \\ 
q_{A\beta} & \epsilon_{AB} e^c 
\end{array}
\right) \, ,
\ee
where the completely antisymmetric tensors in the ${\rm SU(2)}_L$ and ${\rm SU(3)}_C$ spaces are defined 
so that $\epsilon^{12} = 1$ and $\epsilon_{123} = 1$, and
\begin{widetext}
\be
\label{24SMembedd}
(24_{H,F,V})^{\ b}_a = \mathcal{N}_{H,F,V} \\
\left(
\begin{array}{cc}
(O_{H,F,V})^{\ \beta}_\alpha + \tfrac{2}{\sqrt{30}} S_{H,F,V} \delta^{\ \beta}_\alpha & (X_{H,F,V})^{\ B}_\alpha \\
(\overline{X}_{H,F,V})^{\ A}_\beta & (T_{H,F,V})^{\ B}_A - \tfrac{3}{\sqrt{30}} S_{H,F,V} \delta^{\ B}_A
\end{array}
\right) 
\, ,
\ee
\end{widetext}
where we defined the quantities 
\be
\label{defOandT}
O^{\ \beta}_\alpha = \tfrac{1}{\sqrt{2}} O^i (\lambda_i)^{\ \beta}_\alpha 
\quad \text{and} \quad
T^{\ B}_A = \tfrac{1}{\sqrt{2}} T^i (\sigma_i)^{\ B}_A
\ee 
with $\lambda_i$ ($i = 1, \ldots ,8$) and $\sigma_i$ ($i = 1, \ldots ,3$) denoting respectively the Gell-Mann and 
Pauli matrices normalized as $\Tr \lambda_i \lambda_j = 2 \delta_{ij}$ and $\Tr \sigma_i \sigma_j = 2 \delta_{ij}$. 
So, in particular, we have $\Tr T^2 = T^i T^i \equiv \left|T\right|^2$ and 
$\Tr O^2 = O^i O^i \equiv \left|O\right|^2$. 
$\mathcal{N}_{H,F,V}$ is a normalization factor equal to $1$ ($H$) and $\tfrac{1}{\sqrt{2}}$ ($F,V$) respectively.

%===========================================================
%===========================================================
\subsection{Mass spectrum}
\label{massspect}
%===========================================================

The calculation of the tree-level mass spectrum allows to address 
the important question whether the states 
required by the unification pattern can be consistently fine-tuned at the corresponding 
intermediate mass scales. For completeness we report it here, though it can be partially 
found also elsewhere (see for instance Refs.~\cite{Buras:1977yy,Bajc:2006ia,Bajc:2007zf}).

%-------------------------------------------------------------------------
\subsubsection{Scalar sector}
\label{scalsect}
%-------------------------------------------------------------------------

The scalar sector consists in the potential for the GUT-breaking field $24_H$ 
\begin{multline}
\label{V24H}
V_{24_H} = m^2_{24} \Tr 24_H^2 + \mu_{24} \Tr 24_H^3 \\ + \lambda^{(1)}_{24} \Tr 24_H^4 + \lambda^{(2)}_{24} \left( \Tr 24_H^2 \right)^2
\end{multline}
and its interaction with the $5_H$,
\begin{multline}
\label{V5H}
V_{5_H} = m^2_H 5_H^\dag 5_H + \lambda_H ( 5_H^\dag 5_H )^2 + \mu_H 5_H^\dag 24_H 5_H \\ 
+ \alpha \, 5_H^\dag 5_H \Tr 24_H^2 + \beta \, 5_H^\dag 24_H^2 5_H \, .
\end{multline}
SU(5) is spontaneously broken to the SM by 
\be
\label{24HVEV}
\vev{24_H} = \frac{V}{\sqrt{30}} \rm{diag} (2,2,2,-3,-3) \, , 
\ee
where $V$ is a vacuum expectation value in the SM-singlet direction $\vev{S_H}$ (cf.~\eq{24SMembedd}).  
By substituting \eq{24HVEV} into \eq{V24H} the vacuum manifold reads
\begin{equation}
\label{vevV24H}
\vev{V_{24_H}} = m^2_{24} V^2 -\frac{\mu_{24}}{\sqrt{30}} V^3
+ \left( \tfrac{7}{30} \lambda^{(1)}_{24}+\lambda^{(2)}_{24} \right) V^4 
\end{equation}
and the corresponding stationary equation for $V \neq 0$ can be conveniently written as   
\begin{multline}
\label{statcond}
0 = \frac{1}{V}\frac{d\vev{V_{24_H}}}{dV} = \\
2 m^2_{24} -\sqrt{\tfrac{3}{10}} \mu_{24} V
+  \left(\tfrac{14 }{15} \lambda^{(1)}_{24}+4 \lambda^{(2)}_{24}\right) V^2 \, .
\end{multline}
The scalar spectrum is readily obtained by expanding the scalar potential 
around the SM-invariant vacuum configuration in \eq{24HVEV}. 
After trading $m^2_{24}$ by means of the stationary condition in \eq{statcond},
this yields 
\begin{align}
\label{mSH2}
& m^2_{S_H} = - \sqrt{\tfrac{3}{10}} \mu_{24} V 
+ 4 \left( \tfrac{7}{15} \lambda^{(1)}_{24} + 2 \lambda^{(2)}_{24} \right) V^2 
\, , \\
\label{mTH2}
& m^2_{T_H} = -\sqrt{\tfrac{15}{2}} \mu_{24} V 
+ \tfrac{8}{3}\lambda^{(1)}_{24} V^2
\, , \\
\label{mOH2}
& m^2_{O_H} = + \sqrt{\tfrac{15}{2}} \mu_{24} V  
+ \tfrac{2}{3}\lambda^{(1)}_{24} V^2
\, , \\
\label{mXH2}
& m^2_{X_H} = 0
\, , 
\end{align}
with the zero modes corresponding to the would-be Goldstone bosons giving mass to the 
longitudinal components of $X_V$. 
The tree-level vacuum stability (cf.~\eq{vevV24H}) implies 
\be
\label{treelevelvacstab}
\tfrac{7}{30} \lambda^{(1)}_{24} + \lambda^{(2)}_{24} > 0 \, , 
\ee
while, requiring that the scalar masses in \eqs{mSH2}{mOH2} are positive definite (minimum condition) gives 
\be
\lambda^{(1)}_{24} > 0 \, , \quad
-\tfrac{2}{3}\sqrt{\tfrac{2}{15}} \lambda^{(1)}_{24} < \mu_{24}/V < \tfrac{8}{3}\sqrt{\tfrac{2}{15}} \lambda^{(1)}_{24} \, , 
\ee
and
\be
\lambda^{(2)}_{24} > - \tfrac{7}{30} \lambda^{(1)}_{24} + \tfrac{1}{8} \sqrt{\tfrac{3}{10}} \mu_{24}/V \, .
\ee
Since the unification constraints favor a rather light $T_H$ it is interesting 
to work out the vacuum conditions in the limit $m^2_{T_H} \approx 0$. 
In the latter case the heavy spectrum reads
\begin{align}
\label{mSH2lightT}
& m^2_{S_H} \approx 
\left( \tfrac{4}{3} \lambda^{(1)}_{24} + 8 \lambda^{(2)}_{24} \right) V^2 
\, , \\
\label{mOH2lightT}
& m^2_{O_H} \approx \tfrac{10}{3}\lambda^{(1)}_{24} V^2
\, , 
\end{align}
and the absence of tachyons in the scalar spectrum enforces
\be
\label{mincondlightTH}
\lambda^{(1)}_{24} > 0 
\qquad \text{and} \qquad 
\lambda^{(2)}_{24} > -\tfrac{1}{6} \lambda^{(1)}_{24} \, ,
\ee 
which automatically satisfies also the tree-level vacuum stability condition in \eq{treelevelvacstab}.
Notice that a small (positive) value of $\lambda^{(1)}_{24}$ allows 
to consistently keep also the mass of $O_H$ below the GUT scale.  
%In such a case \eq{mincondlightTH} reduces to $\lambda^{(2)}_{24} \gtrsim 0$. 

Finally, by plugging the SM-invariant vacuum configuration of \eq{24HVEV}
into \eq{V5H} we get the spectrum of the fields residing in $5_H$, 
which reads 
\begin{align}
& m^2_{\mathcal{T}} = m^2_H + \sqrt{\tfrac{2}{15}}\mu_H V 
+ \left( \alpha + \tfrac{2}{15} \beta \right) V^2
\, , \\
& m^2_{h} = m^2_H - \sqrt{\tfrac{3}{10}}\mu_H V 
+ \left( \alpha + \tfrac{3}{10} \beta \right) V^2
\, . 
\end{align}
Therefore it is possible to perform the standard doublet-triplet splitting $m^2_{h} \approx 0$, 
which yields in turn
\be
m^2_{\mathcal{T}} \approx \sqrt{\tfrac{5}{6}} \mu_H V - \tfrac{1}{6} \beta V^2 \, .
\ee

\subsubsection{Yukawa sector} 
\label{Yuksect}
On top of the usual Yukawa sector responsible for the masses of the 
charged fermions
\be 
\mathcal{L}_{Ycf} = y_{ij} \overline{5}_F^i 10_F^j 5_H^* 
+ h_{ij} 10_F^i 10_F^j 5_H + \rm{h.c.} + \ldots \, , 
\ee
where the ellipses stand for nonrenormalizable operators needed 
to reproduce the correct mass ratios between down-quarks and 
charged-leptons (see e.g.~\cite{Ellis:1979fg,Dorsner:2006hw}), we add the new Yukawa interactions 
\cite{Bajc:2006ia,Bajc:2007zf}
\begin{multline}
\label{newYuk}
\mathcal{L}_{Y\nu} = y^i_0 \overline{5}_F^i 24_F 5_H 
+ \frac{1}{\Lambda} \overline{5}_F^i 
\left( y^i_1 24_F 24_H  \right. \\ \left. + y^i_2 24_H 24_F + y^i_3 \Tr (24_F 24_H) \right) 5_H + \rm{h.c.} \, ,
\end{multline}
where $\Lambda$ denotes the cutoff of the effective theory. 
After SU(5) breaking \eq{newYuk} yields
\be 
\mathcal{L}_{Y\nu} \ni \ell_i^T (i \sigma_2)^T \left( y^i_T T_F + y^i_S S_F \right) h + \rm{h.c.} \, ,
\ee
where $y^i_T$ and $y^i_S$ are two different linear combinations of $y_0^i$ and $y_a^i V/\Lambda$ ($a=1,2,3$), namely
\begin{align}
& y^i_T = \tfrac{1}{\sqrt{2}} y_0^i - \tfrac{1}{2}\sqrt{\tfrac{3}{5}} (y^i_1 + y^i_2) \frac{ V}{\Lambda} \, , \\
& y^i_S = - \sqrt{\tfrac{3}{10}} y^i_T + \tfrac{1}{\sqrt{2}} y^i_3 \frac{ V}{\Lambda} \, .
\end{align}
In particular, the coupling $y^i_3$ is responsible for the misalignment of the vectors $y^i_T$ and $y^i_S$ in the 
flavour space, thus leading to a rank-2 neutrino mass matrix when integrating out the heavy 
vector-like states $T_F$ and $S_F$: 
\be
\label{mmnu}
m^\nu_{ij} = - \frac{v^2}{2} \left( \frac{y_T^i y_T^j}{m_{T_F}} + \frac{y_S^i y_S^j}{m_{S_F}} \right) \, .
\ee  
Instead, the masses of the new fermions residing in $24_F$ 
are due to the Yukawa-like interactions \cite{Bajc:2006ia,Bajc:2007zf}
\begin{align}
\mathcal{L}_{F} & = m_F \Tr 24_F^2 + \lambda_F \Tr 24_F^2 24_H \nn \\
& + \frac{1}{\Lambda} \left( a_1 \Tr 24_F^2 \Tr 24_H^2 + a_2 \Tr ( 24_F 24_H )^2  \right. \nn \\ 
& + \left. a_3 \Tr 24_F^2 24_H^2 + a_4 \Tr 24_F 24_H 24_F 24_H \right) \, ,
\end{align}
which, after SU(5) breaking, lead to the following spectrum:
\begin{equation}
\label{mSF}
m_{S_F} = m_F - \tfrac{1}{\sqrt{30}} \lambda_F V  
+ \left( a_1 + a_2 + \tfrac{7}{30} (a_3 + a_4) \right) \frac{V^2}{\Lambda}  \, , 
\end{equation}
\be
\label{mTF}
\!\!\!\! m_{T_F} = m_F - \sqrt{\tfrac{3}{10}} \lambda_F V 
+ \left( a_1 + \tfrac{3}{10} (a_3 + a_4) \right) \frac{V^2}{\Lambda}  \, , 
\ee
\be
\label{mOF}
\!\!\!\! m_{O_F} = m_F + \tfrac{2 }{\sqrt{30}} \lambda_F V 
+ \left( a_1 + \tfrac{2}{15} (a_3 + a_4) \right) \frac{V^2}{\Lambda} \, , 
\ee
\be
\label{mXF}
\!\!\!\! m_{X_F} = m_F - \tfrac{1}{2 \sqrt{30}} \lambda_F V 
+ \left( a_1 + \tfrac{13}{60} a_3 - \tfrac{1}{5} a_4 \right) \frac{V^2}{\Lambda}  \, .
\ee
Since unification constraints require a light $T_F$, we must impose $m_{T_F} \approx 0$.  
In turn, the spectrum of the other fields residing in $24_F$ becomes 
\begin{align}
m_{S_F} &\approx \sqrt{\tfrac{2}{15}} \lambda_F V 
+ \left( a_2 - \tfrac{1}{15} (a_3 + a_4) \right) \frac{V^2}{\Lambda} \, , \\
m_{O_F} &\approx \sqrt{\tfrac{5}{6}} \lambda_F V 
- \tfrac{1}{6} (a_3 + a_4) \frac{V^2}{\Lambda} \, , \\
m_{X_F} &\approx \tfrac{1}{2} \sqrt{\tfrac{5}{6}} \lambda_F V 
- \tfrac{1}{2} (\tfrac{1}{6} a_3 + a_4) \frac{V^2}{\Lambda} \, .
\end{align}
Further requiring an intermediate-scale octet ($m_{O_F} \approx 0$), one gets 
\begin{align}
\label{mSFftTO}
m_{S_F} &\approx a_2 \frac{V^2}{\Lambda} \, , \\
\label{mXFftTO}
m_{X_F} &\approx - \tfrac{5}{12} a_4 \frac{V^2}{\Lambda}  \, ,
\end{align}
which shows that the upper bound on the mass of the $X_F$ state is of order $V^2 /\Lambda$.

%-------------------------------------------------------------------------
\subsubsection{Gauge sector}
\label{gaugesect}
%-------------------------------------------------------------------------

The gauge boson masses are obtained from the canonical kinetic term 
\be
\label{cankit}
\tfrac{1}{2} \, \Tr \left( D_\mu \vev{24_H} \right)^\dag D^\mu \vev{24_H}  \, , 
\ee
where $D_\mu$ is the SU(5) covariant derivative
\be
D_\mu 24_H = \partial_\mu 24_H + i g_5 \left[ (24_V)_\mu, 24_H \right] \, .
\ee
After plugging into \eq{cankit} the expression for $\vev{24_H}$ (cf.~\eq{24HVEV}), one finds
\be
\label{gaugespectrum}
m^2_{S_V} = m^2_{T_V} = m^2_{O_V} = 0 \ \ \ \text{and} \ \ \ m^2_{X_V} = \tfrac{5}{12} g_5^2 V^2 \, ,
\ee
leading to the 12 massless modes of the SM gauge bosons, plus the 12 
degrees of freedom of the super-heavy gauge boson $X_V$. 

%===========================================================
\subsection{Low-energy interactions}
\label{lowenint}
%===========================================================

Here we derive the interactions in the low-energy effective theory featuring 
the SM fields and the five intermediate mass-scale states $S_F$, $T_F$, $O_F$, $T_H$ and $O_H$.  

Let us start from the Yukawa-like interactions. At the leading order in $V/\Lambda$ we find
\begin{multline}
\mathcal{L}_F \ni 
y_{STT} \, S_F \, \Tr  T_F T_H \\
+ y_{SOO} \, S_F \, \Tr  O_F O_H 
+ y_{OOO} \, \Tr  O_F^2 O_H \, ,
\end{multline}
with $O_{H,F}$ and $T_{H,F}$ defined in \eq{defOandT} and  
\begin{align} 
& y_{STT} = - \sqrt{\tfrac{3}{10}} \lambda_F 
+  \left( a_2 + \tfrac{3}{5} (a_3 + a_4) \right) \frac{V}{\Lambda} \, , \\
& y_{SOO} = \sqrt{\tfrac{2}{15}} \lambda_F 
+ \left( a_2 + \tfrac{4}{15} (a_3 + a_4) \right) \frac{V}{\Lambda}  \, , \\
& y_{OOO} = \tfrac{1}{2} \lambda_F 
+ \sqrt{\tfrac{2}{15}}  (a_3 + a_4) \frac{V}{\Lambda}  \, .
\end{align}
Notice that the ${\rm SU(2)}_L$ invariant $\Tr T_F^2 T_H$ is zero by antisymmetry. 

The couplings $y_{STT}$, $y_{SOO}$ and $y_{OOO}$ 
have an upper bound of $\mathcal{O} \left( V/\Lambda \right)$, 
since the unification pattern requires a splitting among the masses in~\eqs{mSF}{mXF}.
In particular, for light $T_F$ and $O_F$ they reduce to 
\begin{align} 
& y_{STT} \approx \left( a_2 + \tfrac{1}{2} (a_3 + a_4) \right) \frac{V}{\Lambda} \, , \\
& y_{SOO} \approx \left( a_2 + \tfrac{1}{3} (a_3 + a_4) \right) \frac{V}{\Lambda}  \, , \\
& y_{OOO} \approx  \tfrac{1}{2} \sqrt{\tfrac{5}{6}}  (a_3 + a_4) \frac{V}{\Lambda}  \, .
\end{align}
The other interactions relevant for the scalar sector are
\begin{multline}
V_{24_H} \ni 
\mu_{O} \, \Tr  O_H^3
+ \tfrac{\lambda_{T}}{2} \left( \Tr  T_H^2 \right)^2 \\
+ \lambda_{TO} \, \Tr  T_H^2 \, \Tr O_H^2
+ \lambda_{O} \left( \Tr  O_H^2 \right)^2 \, ,
\end{multline}
where 
\begin{align} 
\label{muO}
& \mu_{O} = \mu_{24} + 4 \sqrt{\tfrac{2}{15}} V \lambda^{(1)}_{24}\, , \\
\label{lamT}
& \lambda_{T} = \lambda^{(1)}_{24} + 2 \lambda^{(2)}_{24} \, , \\
\label{lamTO}
& \lambda_{TO} = 2 \lambda^{(2)}_{24} \, , \\
\label{lamO}
& \lambda_{O} = \tfrac{1}{2} \lambda^{(1)}_{24} + \lambda^{(2)}_{24} \, .
\end{align}
Notice that the ${\rm SU(2)}_L$ invariant $\Tr T_H^3$ is zero by antisymmetry 
and that we also used the relations $\Tr T_H^4 = \tfrac{1}{2} (\Tr T_H^2)^2$ and 
$\Tr O_H^4 = \tfrac{1}{2} (\Tr O_H^2)^2$. 

In particular, in the limit of a light $T_H$ (cf.~\eq{mTH2}), 
we have 
\be
\mu_{O} \approx \tfrac{4}{3} \sqrt{\tfrac{10}{3}} V \lambda^{(1)}_{24} 
\, .
\ee
When also $O_H$ is below the GUT scale, $\lambda^{(1)}_{24} \approx 0$ (cf.~\eq{mOH2lightT}), 
which implies
\be
\lambda_{T} \approx \lambda_{TO} \approx 2 \lambda_{O} 
\, .
\ee
Finally, for the scalar interactions of the SM Higgs doublet, $h$, we
obtain 
\begin{multline}
V_{5_H} \ni \mu_{hT} \, h^\dag T_H h 
+ \lambda_h (h^\dag h)^2 \\
+ \lambda_{hT} \, h^\dag h \, \Tr T_H^2 
+ \lambda_{hO} \, h^\dag h \, \Tr O_H^2 
\, ,
\end{multline}
where 
\begin{align} 
\label{muHT}
& \mu_{hT} = \mu_H - \sqrt{\tfrac{6}{5}} V \beta \, , \\
\label{lambdaH}
& \lambda_{h} = \lambda_{H} \, , \\
\label{lambdaHT}
& \lambda_{hT} = \alpha + \tfrac{1}{2} \beta \, , \\
\label{lambdaHO}
& \lambda_{hO} = \alpha \, ,
\end{align}
and the relation $h^\dag T_H^2 h = \tfrac{1}{2} h^\dag h \, \Tr T_H^2$
has been also employed.

Among the couplings in \eqs{muO}{lamO} and \eqs{muHT}{lambdaHO} only $\lambda_{h}$ is 
fixed in terms of the Higgs boson mass, while the UV constraints coming from the 
SU(5) symmetry reduce only partially the allowed parameter space. 
On the other hand, 
an important constraint for the scalar parameters 
is given by the requirement of perturbativity (cf.~the analysis in \sect{scalarsi}).

%%%%%%%%%%%%%%%%%%%%%%%%%%%%%%%%%%%%%%%%%%%
\section{Further analytical results}
\label{extrares}
%%%%%%%%%%%%%%%%%%%%%%%%%%%%%%%%%%%%%%%%%%%

In this Appendix we present the three-loop 
beta functions and the two-loop matching coefficients obtained by including the contribution of the colour octets.

%===========================================================
\subsection{Octet contribution to the beta functions}
\label{Ocontbf}
%===========================================================

The pure-gauge contribution of the colour octets 
to the gauge coupling beta functions can be 
read from \eqs{deltab1T}{deltab3T} after taking into account the proper substitutions:
\begin{itemize}
 \item $\alpha_2 \leftrightarrow \alpha_3$, $G_L \leftrightarrow G_C$, $R_L \leftrightarrow R_C$, 
 $N_h \rightarrow 0$, $Y_f^2 N_f \leftrightarrow Y_Q^2 N_Q$ and $T_{F,H} \rightarrow O_{F,H}$ in \eq{deltab1T}
\begin{multline}
\label{deltab1O}
 \beta_1^{\rm SM + T+O} = \beta_1^{\rm SM + T}+
 \frac{\alpha _1^2}{\pi ^2} \frac{\alpha _3^2 }{\pi ^2} C_{G_C} 
   C_{R_C} N_{R_C} Y_Q^2 N_Q \\
   \times \left(     
   -\frac{11}{576} {N_{O_F}}   - \frac{23}{2304} {N_{O_H}} \right) \, ,
\end{multline}
\item $\alpha_2 \leftrightarrow \alpha_3$, $G_L \leftrightarrow G_C$, $R_L \leftrightarrow R_C$ 
and $T_{F,H} \rightarrow O_{F,H}$ in \eq{deltab3T}
\begin{multline}
\label{deltab2O}
 \beta_2^{\rm SM + T+O} = \beta_2^{\rm SM + T} +
\frac{\alpha _2^2}{\pi ^2} \frac{\alpha _3^2}{\pi ^2} C_{G_C} C_{R_C} N_{R_C} T_{R_L}  N_q \\
 \times \left(-\frac{11}{576}  {N_{O_F}}
   -\frac{23}{2304} {N_{O_H}}\right) \, ,
\end{multline}
 \item $\alpha_2 \leftrightarrow \alpha_3$, $G_L \leftrightarrow G_C$, $R_L \leftrightarrow R_C$, 
 $N_h \rightarrow 0$, $N_f \leftrightarrow N_Q$ and $T_{F,H} \rightarrow O_{F,H}$ in \eq{deltab2T}
\begin{widetext}
\begin{align}
\label{deltab3O}
 \beta_3^{\rm SM + T+O} &= \beta_3^{\rm SM + T} + 
\frac{\alpha _3^2}{\pi ^2} 
\left\{ C_{G_C}\left(
\frac{1}{6}  {N_{O_F}} + \frac{1}{24} {N_{O_H}} \right)
+ \frac{\alpha _3}{\pi} C_{G_C}^2 \left( \frac{1}{3}  {N_{O_F}} + \frac{7}{48}  {N_{O_H}}
\right) \right. \nn \\
&  
+ \frac{\alpha _3^2}{\pi^2} 
\left[ 
\left( 
   \frac{247}{432} C_{G_C}^3
   -\frac{7}{108} C_{G_C}^2 T_{R_C} N_Q \right. \right. \nn \\
   & \left.
   -\frac{11}{576} C_{G_C} C_{R_C} T_{R_C} N_Q
   -\frac{145}{3456} C_{G_C}^3 {N_{O_F}}
   -\frac{277}{6912} C_{G_C}^3 {N_{O_H}}   
\right)
{N_{O_F}} \nn \\
&  +
\left(
   \frac{2749}{6912} C_{G_C}^3
   -\frac{13}{432} C_{G_C}^2 T_{R_C} N_Q \right. \nn \\
   & \left. \left.\left.
   -\frac{23}{2304} C_{G_C}^2 C_{R_C} T_{R_C} N_Q
   -\frac{145}{13824}  C_{G_C}^3 {N_{O_H}}
\right)
{N_{O_H}}
\right]  
\right\} \, ,
\end{align}
\end{widetext}
\end{itemize}
where we used the abbreviations: $Y_Q^2 N_Q = N_{R_L} Y_q^2 N_q + Y_u^2 N_u + Y_d^2 N_d$ 
and $N_Q = N_{R_L} N_q + N_u + N_d$. 
The numerical values of the beta functions specified to the 
SM+T+O model are obtained by the 
following replacements:
(i) $Y_q = \sqrt{\tfrac{3}{5}}\tfrac{1}{6}$, 
$Y_u = -\sqrt{\tfrac{3}{5}}\tfrac{2}{3}$ and 
$Y_d = \sqrt{\tfrac{3}{5}}\tfrac{1}{3}$, denoting the hypercharges of
the SM quarks in the SU(5) normalization;
(ii) $N_q = N_u = N_d = 3$ and $N_{O_{F,H}}=1$ standing for the number of SM quark generations, 
and colour octets.

%===========================================================
\subsection{Octet contribution to the matching coefficients}
\label{Ocontmatch}
%===========================================================

Considering again only the pure-gauge part, 
the color-octet contribution to the matching coefficient $\zeta_{\alpha_3}$ 
is obtained from \eq{zetaalpha2} after the following substitutions: 
$\alpha_2 \leftrightarrow \alpha_3$, $G_L \leftrightarrow G_C$ and $T_{H,F} \leftrightarrow O_{H,F}$. 
These yield in turn 
\begin{widetext}
\begin{align}
\label{zetaalpha3}
\zeta_{\alpha_3} &=
1 + \frac{\alpha_3}{\pi} C_{G_C} \left( -\frac{1}{6}  \ln{\frac{\mu^2}{m_{O_F}^2}} {N_{O_F}} 
-\frac{1}{24} \ln{\frac{\mu^2}{m_{O_H}^2}} {N_{O_H}} \right) \nn \\
& + \frac{\alpha_3^2}{\pi^2} C_{G_C}^2 \left[ 
\left( 
-\frac{7}{288} 
-\frac{1}{12}  \ln{\frac{\mu^2}{m_{O_F}^2}}
+\frac{1}{36}  \ln^2{\frac{\mu^2}{m_{O_F}^2}} {N_{O_F}}
+ \frac{1}{72} \ln{\frac{\mu^2}{m_{O_F}^2}} \ln{\frac{\mu^2}{m_{O_H}^2}} {N_{O_H}} 
\right) {N_{O_F}} \right. \nn \\
& \left. 
+\left(\frac{37}{576}
-\frac{11}{96} \ln{\frac{\mu^2}{m_{O_H}^2}} 
+\frac{1}{576}  \ln^2{\frac{\mu^2}{m_{O_H}^2}} {N_{O_H}} \right) {N_{O_H}} \right] 
\, ,
\end{align}
\end{widetext}
while there is no contribution to $\zeta_{\alpha_{1,2}}$.
Similarly, by performing the same substitutions above in \eqs{gammaTF}{gammaTH}, one obtains  
the one-loop anomalous dimensions for the running masses $m_{O_{F,H}}$, 
which read explicitly
\begin{align}
\gamma^{(1-\rm{loop})}_{m_{O_F}} &= -\frac{\alpha_3}{\pi} \frac{3}{4} C_{G_C} \, , \\  
\gamma^{(1-\rm{loop})}_{m_{O_H}} &= -\frac{\alpha_3}{\pi} \frac{3}{8} C_{G_C} \, .
\end{align}
The numerical values for our model are obtained by replacing the group invariants given
in \Table{tab:inv} and by setting $N_{O_{F,H}}=1$.

\bibliography{bibliography}
\bibliographystyle{h-physrev5}

\end{document}